\documentclass[]{aa}
\usepackage{graphicx}
\usepackage{natbib}
\bibpunct{(}{)}{;}{a}{}{ }
\begin{document}

\title{Timing the bright X-ray population of the core of M31 with XMM-Newton}
\author{ R. Barnard\inst{1}
    \and U. Kolb\inst{1}
    \and J. P. Osborne\inst{2}
     \and C. A. Haswell\inst{1}}

\offprints{R. Barnard, \email{r.barnard@open.ac.uk}}

\institute{ The Department of Physics and Astronomy, The Open University, Walton Hall, Milton Keynes, MK7 6BT, U.K.
     \and The Department of Physics and Astronomy, The University of Leicester, Leicester, LE1 7RH, U.K.}
\date{Received / Accepted}

\abstract{
All 63  discrete X-ray sources brighter than $L_{\rm X}$=5$\times10^{36}$ erg s$^{-1}$ in any of  four XMM-Newton observations of the core of M31 were surveyed for time variability over time-scales of seconds to thousands of seconds, and for intensity variations between observations. This population is likely to be dominated by low mass X-ray binaries (LMXBs). 
  Analysis of the shapes of power density spectra (PDS) of these sources allows us to determine whether they are accreting at a high or low rate; in the observed frequency range  a broken power law (Type A) PDS indicates a  low accretion rate while a simple power law (Type B) indicates a high accretion rate.  
We obtained the 0.3--10 keV luminosities of the sources by modelling their spectral energy distributions. The luminosity function  for Type A PDS appears to consist  of two populations, which we tentatively classify as neutron star and black hole LMXBs.
 We find that 46 sources are likely  X-ray binaries, 13 with possible black hole primaries. Remarkably,  4 out of  11 LMXB black hole candidates appear persistently bright  over $\sim$20 years; there are no persistently bright black hole LMXBs known in our Galaxy.  The lightcurves of 55 X-ray sources had a probability of variability $>$99.9\% in at least one observation. Also, 57 of the sources show a luminosity variation $\ga$5$\sigma$ between observations; 7 of these are classed as transients, since they are absent in at least one observation, and vary in luminosity by a factor of $\ga$10. Scaling this sample with the known Galactic LMXBs, we find  that the M31 sample has $\sim$50\% of the dippers,  Z-sources and transients. 
}
\titlerunning{Time variability of  point X-ray sources in \object{M31}}
\maketitle

\section{Introduction}

 At 760 kpc \citep{vdb00} the \object{Andromeda Galaxy (M31)} is the nearest spiral galaxy to our own. The X-ray emission of M31 is dominated by point sources, mostly X-ray binaries (XBs). As low mass stars have considerably longer evolution times than high mass stars,  
 low mass X-ray binaries (LMXBs) are expected to dominate the central  region of the galaxy, which is dominated by the old bulge population; high mass X-ray binaries (HMXBs) are more likely to be found in star forming regions such as the spiral arms. 

X-ray binaries exhibit several types of variability on short  time-scales (i.e. milliseconds to hours). The study of such variability in X-ray sources in external galaxies has been limited by the sensitivity of previous observatories, but the three large X-ray telescopes on board XMM-Newton make such a project possible for the first time; the combined effective area of these telescopes is the largest of any X-ray telescope imaging above 2 keV.

Four XMM-Newton observations were  made of the core of \object{M31} between 2000 June and 2002 January, as part of a survey of the whole galaxy. These observations, and other M31 fields,  represent the best opportunity to date for studying time variability of X-ray sources in an external galaxy. 

 The observed X-ray sources in the field of view are likely to be a mixture of supernova remnants, foreground stars, background galaxies and various accretion powered binaries.  The EPIC-pn detector has a 30$\arcmin$$\times$30$\arcmin$ field of view \citep{stru01}, while the two EPIC-MOS detectors have circular fields of view with a 30$\arcmin$ diameter \citep{turn01}, and all three fields of view are $\sim$co-aligned.

The intensity variations on short and long time-scales in the X-ray population of \object{M31} provide vital clues as to the nature of the sources, and  must be accounted for in combination with the information obtained from spectral analysis. 
 Hence a survey  has been undertaken, in search of such variability in the combined  lightcurves of the EPIC MOS and PN detectors on board XMM-Newton. Variability of X-ray sources over 6 month time-scales is discussed by \citet{osb01}.

The core of M31 has been observed many times with Chandra \citep[e.g.][]{K02, kaa02, will04}.
 { The high spatial resolution of the Chandra  images makes them a vital companion to the XMM-Newton survey}; a single point source in an XMM-Newton observation may be resolved by Chandra into multiple sources \citep[see e.g.][]{bko04}. { By combining results from both  observatories, we can obtain more information than from either alone. }

 In this paper, we summarise the known properties of our X-ray sample, and discuss their power density spectra. 
In Sect.~\ref{spec} we review the current understanding of the X-ray spectra of black hole and neutron star LMXBs, while in Sect.~\ref{src} we discuss the sources of time variability we expected to observe in the M31 LMXBs. Most important is Sect.~\ref{da}, which discusses stochastic variability in disc accretion, and is the foundation of our work. In Sect.~\ref{sedpds} we show how combining the energy and timing analysis allows us to further classify the sources. In Sect.~\ref{litreview} we survey the literature on variability in  Galactic LMXBs with known primaries, to support the work outlined in Sect.~\ref{da}. We then summarise the observations in Sect.~\ref{observ} and present the results in Sect.~\ref{varsum}. Finally, we summarise our findings in Sect.~\ref{summary}.

\section{The X-ray spectra of LMXBs}
\label{spec}

Black hole LMXBs are thought to have five spectral states, reviewed comprehensively in \citet{mr03} and \citet{zg04}. In order of increasing luminosity, these are the quiescent, low/hard, intermediate, high/soft, and very high states. The quiescent state is thought to be an extension of the low/hard state to low luminosities. The spectral energy distribution (SED) of the low/hard state is well described by a hard cut off power law, with a spectral index, $\Gamma$,  $\la$1.8 and a break  at 100--200 keV. In contrast, the high/soft state has a soft, thermal SED, described by a multicoloured disc blackbody. The intermediate and very high states are almost indistinguishable apart from their luminosity, and their SEDs may be described by a two component model, consisting of a thermal component and a soft power law  component, with $\Gamma$ $\ge$ 2.4. 

The SEDs of LMXBs with neutron star primaries are different from those of black hole LMXBs. The continuum spectra of the former are described  by two spatially distinct spectral components: a thermal component, and a non-thermal  component. The non-thermal component arises from unsaturated, inverse-Compton scattering of cool photons on hot electrons in an accretion disc corona (ADC); however, the geometry of these emission regions remains controversial. Observations favour a point-like blackbody (i.e. the neutron star), and an extended ADC with a radius of $\sim$10,000--500,000 km \citep[see e.g.][]{cbc95,bbc01,cbc04}. At low luminosities, the SEDs of neutron star LMXBs may be well described by a single Comptonized component \citep{wsp88}, like { those of} the black hole LMXBs. { In contrast to black hole LMXBS, however}, neutron star LMXBs are dominated by the non-thermal component even at high luminosities \citep[e.g.][]{wsp88, bcb03}; the blackbody component contributes $\la$10\% at low luminosities and up to 30--50\% at higher luminosities \citep{cbc01}. Hence, a thermally dominated SED at high luminosity is the signature of a black hole LMXB \citep{dg03,don04}.

\section{Sources of time variability expected in the LMXBs in \object{M31}}
\label{src}
Long term variability in the point X-ray sources of \object{M31} has been well studied in the last 5 years. Two ROSAT PSPC catalogues of \object{M31} point sources were compiled \citep[][ hereafter S97 and S01 respectively]{S97,S01}, with a combined catalogue of 560 point sources. The 1991 survey revealed 15 variable sources and 9 transient sources { when} compared with a previous Einstein catalogue \citep{tf91}, while comparison between the 1992 and 1991 surveys revealed 34 long term variables and 8 possible transients \citep{S01}. The 1999--2001 Chandra observations reveal variability from observation to observation in $\sim$50\% of the X-ray sources \citep[{ hereafter known as K02}]{K02}, while \citet{tru04} report variability in 80\% of globular cluster sources over 10 XMM-Newton and $\sim$40 Chandra observations.

Galactic LMXBs can exhibit a variety of phenomena resulting in luminosity and colour variations over timescales of milliseconds to hours. Two of these phenomena, namely dipping 
and Z-source movement, should be identifiable in the lightcurves of M31 LMXBs. These will be discussed in Sect.~\ref{dip}  and Sect.~\ref{zs} respectively. Additionally, the disc accretion process results in stochastic variability, discussed in Sect.~\ref{da}.

\subsection{Periodic intensity dips due to photo-electric absorption}
\label{dip}

Several high inclination Galactic LMXBs experience periodic modulation of their lightcurves and SEDs, due to obscuration of the X-ray source by material that is raised out of the body of the disc \citep[see e.g.][]{ws82,cbc95}. Photoelectric absorption leads to preferential removal of soft photons, hence hardening the SED. Dipping has been reported in XMM-Newton observations of two LMXBs in the core of M31: XMMU\thinspace J004314.1+410724 \citep[Bo 158, ][]{tru02} and XMMU\thinspace J004308.6+411247 \citep{man04}, on periods of 2.78 and 1.78 hr respectively.
 The dipping behaviour of Bo 158 is particularly interesting, as the depth of dipping is $\sim$82\% in the 2002, January, observation, yet the dip modulation is $\la$10\% in the 2001, June observation \citep{tru02}.
In \citet{bfh05}, we show that the evolution in the dipping behaviour may be due to { warped} disc precession; we modelled the disc with three dimensional smoothed-particle hydrodynamics, and find that the disc is indeed expected to precess.

\subsection{Z-source movement}
\label{zs}

The Galactic neutron star LMXBs are divided into Z-sources and atoll sources, based on their colour-colour diagrams, or colour-intensity diagrams, and timing properties \citep{hv89}. The six Galactic Z-sources all have luminosities in excess of 10$^{38}$ erg s$^{-1}$, and trace out a three-branched ``Z'' in their colour-colour diagrams. The three branches are called the horizontal branch, normal branch  and flaring branch, and all Galactic Z-sources exhibit a $\sim$6 Hz { quasi-periodic oscillation (QPO)} on the normal branch \citep{hv89}. Branch movement of Z-sources results in significant changes in the SED and luminosity over time-scales of minutes to hours \citep[e.g.][]{bcb03}.  Atoll sources have now been shown to { also}  exhibit Z-shaped colour-colour diagrams \citep{gd02}. However, atoll sources trace their Z-tracks over timescales of $\sim$months, requiring intensity variations  by a factor of $\sim$80, whereas Z-sources only vary in intensity by a factor of $\sim$2--3, and traverse the Z over a few days \citep{mrc02}. The difference between Z-sources and atoll sources is therefore real, but of unknown origin. While most LMXBs have orbital periods $\la$10 hr, the Z-sources have orbital periods $\ga$20 hr \citep[e.g.][]{lvv01}; hence, the secondaries in Z-source are likely to be larger and/or more evolved. It is believed that the difference between Z-sources and atoll sources may be due to some vital difference in their secondaries.

\begin{table*}[!t]
\centering
\caption{ Classifications of the X-ray population of M31 by power density spectrum (PDS) and spectral energy distribution (SED). The SED is classified by the photon index of the power law component, $\Gamma$, or by the temperature of the thermal component, k$T_{BB}$. }\label{class}
\begin{tabular}{lllll}
\noalign{\smallskip}
\hline
\noalign{\smallskip}

PDS & SED & Classification\\
\noalign{\smallskip}
\hline
\noalign{\smallskip}
Type A & $\Gamma$ $\sim$1.5--2.1 & NS or BH LMXB in low/hard state \\
Type A & $\Gamma$ $\sim$2.4--3.0 & BH LMXB in very high/\\
& & intermediate (steep power law) state\\
Type B & BB dominated & BH LMXB in high/soft state\\
& k$T_{\rm BB}$ $\sim$0.7--2 keV  \\
Type B & Not BB dominated& NS LMXB in high state\\ 
\noalign{\smallskip}
\hline
\noalign{\smallskip}

\end{tabular}
\end{table*}

In \citet{bko03}, we identified a likely Z-source in M31, RX\thinspace J0042.6+4115. We produced a colour-intensity plot for four XMM-Newton observations of the source, defining the colour as the intensity ratio of the 4.0--10 keV and 0.3--2.0 keV bands, and found that the data clustered in three regions, roughly describing a Z. The 0.3--10 keV SED of  RX\thinspace J0042.6+4115 required a two component spectral model, consisting of a blackbody component and a power law component arising from Compton scattering, with an unabsorbed luminosity of $\sim$5$\times$10$^{38}$ erg s$^{-1}$ in the 0.3--10 keV band;  RX\thinspace J0042.6+4115 is a good M31 analogue for the Galactic Z-sources \citep{bko03}.

\subsection{Stochastic variability in disc accretion}
\label{da}

Van der Klis (1994) showed that the power density spectra (PDS) of  low mass X-ray
binaries (LMXBs) depend more on the accretion rate than on the nature of the primary. At low accretion rates, the {  fractional} rms variability is high {  \citep[a few tens of percent, ][]{vdk95}} and the PDS  are well described  by a broken power law that changes in spectral index, $\gamma$, from $\sim$0 to $\sim$1 at frequencies higher than a certain break frequency; the break occurs at 0.01--1 Hz \citep{vdk94}.  At higher accretion rates, the
rms variability is only a few percent, and the PDS are characterised by a power law with $\gamma$ $\sim$1--1.5.  Van der Klis (1994) proposed that LMXBs switch from low to high accretion rate  behaviour at some constant  fraction of the Eddington limit, and suggested a transition at 1\% Eddington.  Since we cannot observe the accretion rate, $\dot{M}$, directly, we must use the luminosity to trace the evolution of the PDS with $\dot{M}$. While the luminosity is a good indicator of $\dot{M}$, the correspondence is not simple \citep[see e.g.][]{don04}. We define $l_{\rm c}$ as 
\begin{equation}
l_{\rm c} = \frac{L_{\rm c}}{L_{\rm Edd}},
\end{equation}
where $L_{\rm c}$ is the luminosity of transition 
and $L_{\rm Edd}$ is the Eddington limit.

We denote the low accretion rate variability (broken power law PDS) as Type A and high accretion rate variability (simple power law PDS) as Type B \citep*{bko04}. We realised that if the transition from Type A to Type B occurred at some constant  $l_{\rm c}$, then black hole LMXBs would exhibit Type A variability at higher luminosities than neutron star LMXBs \citep{bok03,bko04}. We have some evidence that $l_{\rm c}$ $\sim$0.1 in the 0.3--10 keV band, from Galactic LMXBs with known primaries, and the globular cluster X-ray sources in M31. We present results from the literature in Sect.~\ref{litreview}, and evidence from the M31 globular cluster X-ray population in Sect.~\ref{globclust}.

\section{Combining SED and Timing analysis}
\label{sedpds}
  
\citet{mr03} recently reviewed the spectral and timing properties of Galactic black hole binaries { in} their different states. Ten years after the \citet{vdk94} paper, the general scheme of black hole behaviour still holds. For the hard (low/hard) state, \citet{mr03} observe the SED to be dominated by a power law component with $\Gamma$ $\sim$1.5--2.1, while { the} PDS  is of Type A, with r.m.s. variability $\sim$0.1--0.3 in the 0.1--10 Hz band. 
For the thermally dominated (high/soft) state, a disc black body with temperature ~0.7--2 keV contributes  $\ga$90\% of the flux  in the 0.3--10 keV band; Type B PDS are observed with fractional  r.m.s. variabilities of $\sim$0.01--0.06 over the 0.1--10 Hz frequency range. \citet{mr03} replace ``very high state'' and ``intermediate state''  with the ``steep power law'' state. The SED contains a power law component with $\Gamma$ $\sim$2.4--3.0 which contributes $>$20\% of the 2--20 keV emission. If the power law component contributes 20--50\%, then a QPO is seen in the PDS at $\sim$0.1--30 Hz; however, if the power law component contributes $>$50\%, then no QPO is seen.  The PDS are of Type A with break frequencies $\sim$0.1--10 Hz, i.e. at higher frequencies than in the low hard state.  There appears to be no consensus on the r.m.s. variability of the PDS associated with the very high/intermediate state; however, \citet{vdk95} states that the r.m.s. variability lies between those of the low/hard and high/soft states.

Joint analysis of the SEDs and PDS of the M31 X-ray sources therefore allows  greatly improved  classification of the population. In Table ~\ref{class}, we summarise our classification { scheme}.

\section{PDS transitions in Galactic LMXBs with known primaries}
\label{litreview}

We have performed a literature survey of Galactic LMXBs, with the goal of constraining $l_{\rm c}$. We considered LMXBs with known primaries that have either shown transitions from Type A to Type B,  exhibited Types A and B  within a narrow luminosity range, or have peaked during outburst without making the transition to Type B. The motivation is to discover whether the transitions occur at some constant fraction of the Eddington limit; if true, then the transition luminosity is a probe of the primary mass. Since we are unable to measure the bolometric luminosities of M31 X-ray sources, we have converted all luminosities presented in the literature to luminosities in the 0.3--10 keV band. 

\begin{table*}
\caption{Properties of five transient black hole LMXBs  that exhibited hard outbursts.   }\label{hxts}
\begin{centering}
\begin{tabular}{cccc}
\noalign{\smallskip}
\hline
\noalign{\smallskip}
Name &  $L_{\rm  A}$$^a$  & Mass / M$_{\sun}$&  Min $l_{\rm c}$$^b$\\
\noalign{\smallskip}
\hline
\noalign{\smallskip}
{\object GS\thinspace 2023+338} & $>$30 (2--37 keV) $^c$ &  $\sim$12$^d$ & 0.06$\pm$0.01$^e$\\
{\object GRS\thinspace J1719$-$24} & 3 (3--200 keV)$^f$ & $\sim$5$^f$ & 0.06$\pm$0.01$^g$\\
{\object GRS\thinspace 1737$-$31} & 3 (2--200 keV)$^h$ & 4--14 & 0.015$\pm$0.005$^i$\\
{\object GRO\thinspace 0422+32} & 6.4$\pm$0.8 (3--200 keV)$^j$& 4.0$\pm$1.0$^k$ & 0.10$\pm$0.03$^l$\\
{\object XTE J1118+480} & 0.04 (2--12 keV)$^m$ & 6.0--7.7$^n$ & 0.00067$\pm$0.00018$^o$\\
\noalign{\smallskip}
\hline
\noalign{\smallskip}
\end{tabular}
\end{centering}
\\$^a$ Highest luminosity where Type A PDS is observed / 10$^{37}$ erg s$^{-1}$;
$^b$minimum $l_{\rm c}$ in 0.3--10 keV band;
$^c$\citep[][ and references within]{miy92,oost97}; $^{d}$\citet{shab94}; $^e$\citep[assuming power law X-ray spectrum with spectral index  $\alpha$ = 1--1.4][]{miy92}; $^f$\citet{rev98}, assuming distance of 2.8 kpc \citep{dv94};
$^g$ $\alpha$ = 2--2.3 \citep{vdh96}; 
$^h$\citet{cui97};
$^i$$\alpha$ $\sim$1.7 \citep{cui97};
$^j$obtained by comparison of 40--150 keV flux with GRS\thinspace 1737$-$31, which has a similar $\alpha$ \citep{vdh99} and distance \citep{gh03};
$^k$\citet{gh03};
$^l$$\alpha$ $\sim$2.1 \citep{vdh99};
$^m${Flux = 42 mCrab \citep{hh03}, distance=1.71$\pm$0.05 kpc \citep{chat03}};
$^n${\citet{wag01}};
$^o$$\alpha$ = 1.78 \citep{mh01}
\end{table*}

\subsection{ $l_{\rm c}$ for a Galactic neutron star LMXB}

4U\thinspace 1705$-$44 is a Galactic LMXB that exhibits X-ray bursts, and hence contains a neutron star \citep{lang87}. It exhibited a Type A  PDS in the faintest of four EXOSAT observations, and a Type B PDS in the next faintest;  the respective 1--20 keV fluxes were $\sim$1.3$\times$10$^{-9}$ and $\sim$1.8$\times$10$^{-9}$ erg cm$^{-2}$ s$^{-1}$ \citep{lang87,lang89}. We obtained 0.3--10 keV fluxes for these observations using the best fit spectral models obtained by \citet{lang87}, yielding fluxes of 1.2 and 1.9 $\times$10$^{-9}$ erg cm$^{-2}$ s$^{-1}$ respectively. Hence, an accurate distance { could} yield a tight constraint on $l_{\rm c}$. The distance to 4U\thinspace 1705$-$44 has been estimated using X-ray bursts as standard candles \citep[see][]{kul03};  \citet{cs97} find a distance of 11 kpc from Einstein data, while \citet{cki02} obtain a distance of 8.6 kpc using BeppoSAX. If we assume that the distance lies between these two values,  then $l_{\rm c}$ is consistent with  ~0.10$\pm$0.04.

\subsection{ $l_{\rm c}$ for  Galactic black hole LMXBs}

Of the 18 confirmed Galactic black hole X-ray binaries, 3 are high mass X-ray binaries (HMXBs), and are persistently bright, and the rest are transient LMXBs.   In addition, there are a further 22 transient LMXBs that are black hole candidates \citep{mr03}. These 37 LMXB black hole or black hole candidate transients make up $\sim$50\% of the known Galactic LMXB transients \citep[see e.g.][]{int04}.  In general  outbursts last several months and the X-ray luminosity can increase by a factor of 10$^7$;   outbursts are on average separated by years of quiescence \citep{cst97,int04}. 

  Transient LMXBs are hysteretic, in that  the hard to soft transition during the rise of the  outburst occurs at a higher luminosity than the soft to hard transition during decay \citep{miy95,mac03}. We were interested in obtaining the maximum value for $l_{\rm c}$, as this would be most sensitive to the Eddington limit.  Hence, estimates of $l_{\rm c}$ in black hole LMXBs were restricted to those that had either been observed during the rise of the outburst, or exhibited outbursts where the transition from low/hard state to high/soft state was not made \citep[see][ for a review]{bbf04}. Unfortunately, most X-ray observations of Galactic black hole LMXBs during outburst have been during the decay phase, where the spectral transition occurs at $l$ $<$ $l_{\rm c}$ due to hysteresis, and are hence unsuitable for this work.

Nine  black hole LMXBs  have exhibited outbursts where they remained in the low/hard state \citep[for a review see][]{bbf04}. Five of these have published   Type A PDS, distances and mass estimates; we have used published results to obtain the   corresponding minimum value of $l_{\rm c}$ in the 0.3--10 keV band. The results are presented in Table~\ref{hxts}; the mass of GRS\thinspace 1737$-$31 is given by the mass range of known black holes. The  minimum $l_{\rm c}$ values in Table~\ref{hxts} cover the range 7$\times10^{-4}$--0.10, with $l_{\rm c}$ $\ga$0.06 for 3 out of 5 sources.  It is reasonable to conclude that  $l_{\rm c}$ $\ga$0.05 in the 0.3--10 keV band.

  Two of the    confirmed black hole LMXBs,  {\object GX\thinspace 339$-$4} \citep{zdz04} and \\{\object XTE\thinspace J1550$-$564} \citep{rod03}, have recently been caught during the rise of the outburst by the RXTE-ASM, allowing monitoring of the entire outburst by the main instruments of RXTE. \citet{zdz04} present results from two outbursts of GX\thinspace 339$-$4 that were observed with RXTE, and a lightcurve spanning 1987--2004, covering $\sim$15 outbursts. They found that the  luminosity of the transition between spectral states during the rise of the outburst depended on the history of the disc: the transition during the first outburst observed by RXTE  was at a bolometric luminosity, $L_{\rm bol}$, of 7\% Eddington, while the transition during the rise of the second outburst was at $L_{\rm bol}$ $\sim$ 20\% Eddington; these correspond to $l_{\rm c}^{0.3-10 keV}$ $\sim$ 0.04 and 0.10 respectively. Prior to the first outburst, GX\thinspace 339$-$4 was quasi-persistently bright with frequent intensity dips, whereas the second outburst was preceded by a $\sim$1000 day low state \citep{zdz04}. The second outburst is more representative of a canonical outburst, and has similar characteristics to the April--June 2000 outburst of XTE\thinspace J1550$-$564, where the spectral state transition during the rise also occurred at $L_{\rm bol}$ $\sim$20\% Eddington, corresponding to $\sim$10\% in the 0.3--10 keV band. The spectral transition was not observed in its 1998 outburst, instead the very high, high and intermediate states were observed \citep{sob99}.  We note, however, that \citet{zdz04} and \citet{rod03} identified the transitions from the changing SEDs, not their PDS. We assume that the transition from Type A to Type B variability is simultaneous with the transition from the low/ hard to the high/soft state, although this may not be the case.

\subsection{An empirical value for $l_{\rm c}$}
 Our considerations show that eight Galactic LMXBs with known primaries (7 black holes, 1 neutron star) are consistent with $l_{\rm c}$ $\sim$0.1 in the 0.3--10 keV band. It is beyond the scope of this paper to conduct a comprehensive survey of PDS in Galactic LMXBs. Instead, we intend these results to be  a useful starting point, prior to a thorough investigation of PDS transitions in Galactic X-ray binaries with known primaries.

\citet{es97} modelled the hard to soft transition using observations of the black hole transient Nova Muscae 1991. They found that the transition occurs at a critical accretion rate as a fraction of the Eddington limit, $\dot{m}_{\rm crit}$ that depends on the disc viscosity $\alpha$: $\dot{m}_{\rm crit}$ $\sim$1.3 $\alpha^2$. They assume $\alpha$ = 0.25, giving $\dot{m}_{\rm crit}$ $\sim$0.08. \citet{es98} then successfully applied the model to \object{Cygnus\thinspace X-1}, GRO\thinspace J0422+32 and GRO\thinspace J1719$-$24, obtaining $\dot{m}_{\rm crit}$ $\sim$0.1. However, it is uncertain whether $l_{\rm c}$ $\equiv$ $\dot{m}_{\rm crit}$. 

 We note that    \citet{es98} find that the SED of the  GRO\thinspace 0422+32 at the peak of its outburst is consistent with an accretion rate of $\sim \dot{m}_{\rm crit}$; this is interesting because we find  the 0.3--10 keV luminosity of GRO\thinspace 0422+32 at that time to be 10$\pm$3\% Eddington. Results from the other six black hole LMXBs are also consistent with  $l_{\rm c}$ $\sim$0.1 in the 0.3--10 keV  band, although they cannot constrain $l_{\rm c}$.  For the remainder of the paper, we assume that $l_{\rm c}$ = 0.1 for all LMXBs.

The maximum mass for a neutron star is generally accepted to be 3.1 M$_{\odot}$ \citep[e.g.][]{kk78}, but could be as high as 6 M$_{\odot}$, depending on the equation of state \citep{sri02}. We must appeal to observations: all measured neutron star masses are less than 2 M$_{\odot}$, and most are consistent with 1.35$\pm$0.04 M$_{\odot}$ \citep[ and references within]{sri02}.  We will assume a maximum neutron star mass of 3.1 $M_{\odot}$. This allows us to postulate a maximum luminosity for Type A variability in a neutron star LMXB of  $\sim$4$\times$10$^{37}$ erg s$^{-1}$ in the 0.3--10 keV band.

\begin{table}[!t]
\centering
\caption{Journal of XMM-Newton observations of the \object{M31} core. { The observation number, date, total exposure (Exp), exposure of the good time interval (GT) and filter are shown.}}\label{journ}
\begin{tabular}{lllll}
\noalign{\smallskip}
\hline
\noalign{\smallskip}

Observation & Date & Exp& GT  & Filter\\
\noalign{\smallskip}
\hline
\noalign{\smallskip}
1 &  25/07/00 & 34 ks& 27 ks& Medium \\
2 & 27/12/00 & 13 ks& 13 ks& Medium\\
3 & 29/06/01 & 56 ks & 25 ks&Medium \\
4 & 06/01/02 & 64 ks& 64 ks& Thin\\
\noalign{\smallskip}
\hline
\noalign{\smallskip}

\end{tabular}
\end{table}

\section{ The Observations}
\label{observ}
Four observations of the core of \object{M31} were made with XMM-Newton, separated by 6 months; a journal of the observations is presented in Table~\ref{journ}. Observations 1 and 3 suffered intervals of background flaring; for each of these observations, a single flare-free  good time interval was used. While previous surveys used many short pointed observations, the four XMM-Newton observations were long and uninterrupted; together with the unprecedented effective area of XMM, the observations yielded up to 40 times the counts of the best previous  observations.

Data reduction was performed using the XMM-Newton SAS version 6.0.0, and the products were analysed with XANADU and the FTOOLS software suite.
 We conducted a timing survey of the   63 brightest X-ray sources in these  four XMM-Newton observations of the central region of M31, extracting the  PDS and 0.3--10 keV luminosity of each source, assuming a distance  to M31 of 760 kpc \citep{vdb00}.

We selected our sources by intensity; the mean  intensity of the background subtracted, 0.3--10 keV pn lightcurve had to exceed 0.02 count s$^{-1}$ in at least one observation; this ensured at least $\sim$250 source counts in the pn spectrum from the shortest observation.  For each source, data were extracted from a circular region that was centred on the source; those sources within 8$\arcmin$ of the nucleus had extraction regions with 20$\arcsec$ radii, due to crowding. Sources further than 8$\arcmin$ from the nucleus had extraction regions of 40$\arcsec$. A background region was chosen for each source to be a source-free region with the same size as the source region, on the same CCD and at a similar off-axis angle. 

Source and background lightcurves with 2.6 s time resolution were obtained for each source  from the three EPIC detectors: MOS1, MOS2 and pn. The background-subtracted lightcurves were corrected for vignetting and summed, giving a combined EPIC lightcurve for every source, for each of the four XMM-Newton observations. PDS were then obtained for each lightcurve using the {\sc powspec} program in  FTOOLS, averaging over intervals of 333 s  and 666 s with time bins of 5.2 s (intervals of 64 and 128 bins  respectively). PDS were also made for the combined EPIC background lightcurves. This allowed us to determine whether the variability observed in the source PDS was { caused} by variation in the background.

The output PDS were then fitted with power law and broken power law models, using custom software. The power law model was parameterised by the spectral  index, $\gamma$, and normalisation, while the broken power law model was characterised by the  spectral index changing from $\alpha$ to $\beta$ at the break frequency $\nu_{\rm c}$, normalised by the power at $\nu_{\rm c}$.  
 To ensure that the broken power law fits described Type A PDS, we set $\alpha$ = 0. We then conducted an F-test on the best fit power law and broken power law models, to determine whether the broken power law gave a significantly better fit.

 \begin{table*}[!ht]   
\renewcommand{\baselinestretch}{1}
\caption{\footnotesize Coordinates, corresponding catalogue number in S97, S01,  K02 and W04 catalogues, along with properties recorded in these surveys, PDS types observed and the most likely primary when known. The codes in the properties column are explained in the text. { References for previously known properties are given in the foot notes;  a property without a reference is first reported in this work.} }\label{prop}
\begin{tabular}{llllllllllllll}
\noalign{\smallskip}
\hline
\noalign{\smallskip}
Source & R.A. & Dec. & S97 & S01 & K02/W04 &Properties & PDS & Primary \\
\noalign{\smallskip\hrule\smallskip}
S1   & 00:42:05.71 & +41:13:29.7$^{j}$ &&& r3-125 & $T^{c}$ $v$  & $A$ $-$  \\
S2  & 00:42:07.619 & +41:18:15.17$^{d}$ &  141& 139 & r3-61& $t^{a}$&  $-$   \\
S3  & 00:42:08.952 & +41:20:48.42$^{d}$ &142  & 142 & r3-60&  $v$&  $A$ $-$  \\
S4 & 00:42:09.372 & +41:17:45.63$^{d}$ & &&r3-59 & $glob^{d}$ $t^{l}$ $v$ & $B$ $-$ \\
S5  & 00:42:12.026& +41:17:58.86$^{d}$ & && r3-54& $t^{c}$ $glob^{d}$ $v$&   \\
S6  & 00:42:13.017 & +41:18:36.73$^{d}$ &146  & 144 & r3-52 &$t^{a}$ $fg^{a}$  $v$&  $-$\\
S7  & 00:42:15.011& +41:12:34.23$^{d}$ &&& r3-50 & $t^{d}$  $v$&  $A$ $B$ & $bh$\\
S8  & 00:42:15.571 & +41:17:21.11$^{d}$ & 151 & &r3-47 & $t^{a}$ $v$ &  $A$  \\ 
S9 & 00:42:15.86& +41:01:14.7$^{j}$& 150 & 147 & s1-7 & $t^{a}$ $glob^{a}$&  $-$\\
S10  & 00:42:18.241 & +41:12:23.53$^{d}$ &155  & 153 &r3-45&  $t^{a}$ $v$&  $-$   \\
S11 & 00:42:18.534 & +41:14:01.69$^{d}$ & 158 & 154 & r3-44& $t^{a}$ $glob^{a}$ $v$ & $-$ \\
S12 & 00:42:21.382& +41:16:01.32$^{d}$ & &&r3-42 & $t^{d}$ $v$ &  $A$  & $bh$ \\
S13  & 00:42:22.316&+41:13:33.99$^{d}$&&&r3-40 & $t^{d}$  $v$ & $A$  \\
S14 & 00:42:22.841 & +41:15:35.14$^{d}$ &163  & 159 &r3-39 &   $t^{d}$ $v^{e}$&    $A^{g}$  & $bh^{g}$ \\
S15 & 00:42:25.941 & +41:19:15.27$^{d}$&  & & r2-36 &  $glob^{d}$ $t^{d}$ $v$& $B$ \\
S16  & 00:42:26.047 & +41:25:52.74$^{d}$ & 167 & 166& r3-87 & $v$& $A$  $B$ \\ 
S17 & 00:42:28.089& +41:09:59.84$^{d}$ & & & r3-36 & $t^{d}$ $v$ & $A$ $-$  \\
S18 & 00:42:28.190 & +41:12:22.76$^{d}$ & 172 & 169 & r2-35 & $t^{a}$ $v$ &     \\
S19 & 00:42:28.789& +41:04:34.98$^{d}$&173 & 170& r3-111& $t^{a}$ $v$ & $B$   \\
S20 & 00:42:31.045 & +41:16:21.74$^{d}$ & 176 & 174 & r2-34&    $t^{d}$ &  $-$ &  \\
S21 & 00:42:31.154 & +41:19:39.19$^{d}$ &175  & 175 & r2-33&  $glob^{a}$ $t^{d}$   $v$ & $A$ $-$    \\
S22 & 00:42:31.979 & +41:13:14.24$^{d}$ & 177& 177  & r2-32 &  $t^{d}$  $m(2)$ $v$  &$B$  \\ 
S23 & 00:42:34.361 & +41:18:09.60$^{d}$&&&r2-29 &  $T^{c}$ $v$ &$A$  \\
S24 & 00:42:35.121 & +41:20:06.09$^{d}$ & 182 & & r2-27 &  $t^{d}$  $v$& $A$ $-$ \\
S25 & 00:42:38.503 & +41:16:03.80$^{d}$ & 184& 184 & r2-26 &   $t^{d}$ $zs^{h}$   $v$& $B$ $-$ & $ns$ \\ 
S26  & 00:42:39.451 & +41:14:28.52$^{d}$ & &&r2-25 &$t^{d}$ $v$& $-$ \\
S27  & 00:42:40.121 & +41:18:45.38$^{d}$ &&& r2-24 & $t^{d}$ $v$ &$A$  \\
S28 & 00:42:41.566 & +41:21:06.02$^{d}$ &&&r3-31 & $t^{d}$ $v$& $A$  \\
S29  & 00:42:43.225 & +41:13:19.48$^{d}$ & & & r2-19& $v$& $A$ $-$  \\
S30 & 00:42:44.766 &  +41:11:37.76$^{d}$ & 195 & 194 &r3-29 &   $t^{d}$ $v$&   $B$ &$ns$  \\
S31 & 00:42:47.089 & +41:16:28.65$^{d}$ & 198& 195 & r1-2 &  $t^{c}$  $pn^{d}$  $m(4)$ $v$ &$A$  \\
S32  & 00:42:48.450 & +41:25:23.10$^{d}$ &  201& 198 & r3-25 & $t^{d}$ $v$  &$B$ $-$ \\ 
\noalign{\smallskip}
\hline
\noalign{\smallskip}
\end{tabular}
\\$^{a}$S97, S01; $^{b}$\citet{kah99}; $^{c}$\citet{osb01};  $^{d}$K02; $^{e}$\citet{kaa02}; $^{f}$\citet{tru02}; $^{g}$\citet{bok03}; $^{h}$\citet*{bko03}; $^{i}$\citet{man04}; $^{j}$W04; $^{k}$\citet*{bko04}; $^{l}$\citet{tru04}

\end{table*}

\setcounter{table}{3}
 \begin{table*}[!t]   
\renewcommand{\baselinestretch}{1}
\caption{ continued
  }
\begin{tabular}{lllllllllllllllllllll}
\noalign{\smallskip}
\hline
\noalign{\smallskip}
Source & R.A. & Dec. &  S97 & S01 & K02/W04 &Properties & PDS & Primary\\
\noalign{\smallskip\hrule\smallskip}

S33 & 00:42:48.52 & +41:15:21.4$^{j}$ &200  & 197 &r1-1&  $t$ $m(2)$  $v$ & $A$ $B$ $-$ \\
S34 & 00:42:52.450 & +41:18:54.75$^{d}$ &206  & 200 & r2-13 &   $t^{d}$ $v^{e}$   &$B$ $-$  \\
S35 & 00:42:52.450 & +41:15:40.20$^{d}$ & 208 & 203  & r2-12 &  $t^{a}$  $sss^{b}$ $v$ & $-$\\
S36 & 00:42:54.3 & +41:30:49  &&&&& $-$  \\
S37 & 00:42:54.859 & +41:16:03.46$^{d}$ & 210  & & r2-11 &    $v$ $t^{d}$   $r?$ &$A$ $-$ & $bh$\\
S38 & 00:42:55.316 & +41:25:56.60$^{d}$ & 211 & 206 &r3-23 & $snr^{a}$  $t$ $v$  & $A$  \\
S39 & 00:42:55.550 & +41:18:35.44$^{d}$&&&r2-9 & $t^{d}$ $m(2)$ & $A$ $-$  \\
S40 & 00:42:57.854 & +41:11:04.59$^{d}$&213 & 209 &r3-22&  $v$   &$A$ $-$ & $bh$ \\
S41 & 00:42:58.257 & +41:15:29.46 $^{d}$&&&r2-7 & $t^{d}$  $v$&  $A$ \\
S42 & 00:42:59.594 & +41:19:19.72$^{d}$ &217  & 210 &r2-6 &  $glob^{a}$  $t^{d}$$-$   \\
S43 & 00:42:59.803 & +41:16:06.01$^{d}$ &&&r2-5 & $t^{d}$ $glob^{d}$ $v$& $A$ $-$  \\
S44 & 00:43:00.87 & +41:30:07.7$^{j}$&218  &212&n1-76 &$glob^{a}$& $-$ \\
S45 & 00:43:02.958 & +41:20:42.54$^{d}$ &&& r3-21 & $pn^{d}$ $t$ $v$&  $A$ $B$ \\
S46 & 00:43:03.089 & +41:10:15.18$^{d}$ & 211 & & r3-20 & $t$ $v$ &$A$ $B$  \\
S47 & 00:43:03.163 & +41:15:28.00$^{d}$ &  226& 214 & r2-3 &  $glob^{a}$ $T^{d}$  $m(2)^{k}$  &  $A^{k}$ $-$& $bh^{k}$ \\
S48 & 00:43:03.231  & +41:21:22.42$^{d}$ &222  & 216 & r3-19 &  $glob^{a}$ $t^{d}$  $v$ & $A$ \\
S49 & 00:43:03.812 &+41:18:05.23$^{d}$ & 223& 217 &r2-2 &    $glob^{a}$ $t^{d}$ $v$ & $A$   \\
S50 & 00:43:04.186 & +41:16:01.62$^{d}$ & & & r2-1 & $t^{d}$ $v$ & $A$ $B$  \\
S51  & 00:43:05.66& +41:17:03.3$^{j}$ &&&r2-69 &  $T^{j}$   $v$& $A$ & $bh^{j}$  \\
S52 & 00:43:08.63 & +41:12:50.1$^{j}$ & & & r3-17 & $v^{i}$ $dip^{i}$ $t$ & $A$ $B$ $-$ \\ 
 S53 & 00:43:09.791 & +41:19:01.22$^{d}$ & 226 & 222 & r3-16 & $t^{a}$ $T^{d}$ $r$ $v$ & $A$ & $bh$  \\
 S54 &  00:43:10.587 & +41:14:51.55$^{d}$ &  228& 223 &r3-15 & $glob^{a}$ $t^{d}$   $v$ &   $-$&    \\
S55 & 00:43:14.245 & +41:07:25.42$^{d}$ & 229 & 227 & r3-112 & $glob^{a}$ $dip^{f}$ $t^{f}$ $v$ &$B$ \\
S56 & 00:43:18.773 & +41:20:18.52$^{d}$& 235 & 235 & r3-8& $t^{a}$  $sss^{c}$ $v^{e}$  & $A$ & \\
S57 & 00:43:19.52 & +41:17:56.7$^{j}$& && r3-126& $sss^{c}$ $pulse^{c}$ $T^{c}$  $v$ $m(2)$ & $A$  \\
S58 & 00:43:27.763 & +41:18:29.83$^{d}$ & 240 & 241 & r3-63 & $snr^{a}$ $t^{a}$ $sss^{e}$ $v$  &$A$   \\
S59 & 00:43:29.038 & +41:07:49.11$^{d}$ & 241 & 242 & r3-103& $t$  $v$ & $A$ $-$&  $bh$ \\
S60 & 00:43:32.382 & +41:10:41.66$^{d}$ & 243 & 244 & r3-3 & $t$  $v$ &  $A$ & $bh$ \\
S61 & 00:43:34.332 & +41:13:23.72$^{d}$ & 244 & 247 & r3-2 & $t$ $v$ &  $A$  \\
S62 & 00:43:37.191 & +41:14:407$^{d}$ & 247 & 249 & r3-1 & $glob^{a}$ $t$  $v$&   &    \\
S63 & 00:43:54.2 & +41:16:44$^{c}$ & 256 & 262 & n1-79& $t$ $v$ & $A$ & $bh$\\
\noalign{\smallskip}
\hline
\noalign{\smallskip}
\end{tabular}
\\$^{a}$S97, S01; $^{b}$\citet{kah99}; $^{c}$\citet{osb01};  $^{d}$K02; $^{e}$\citet{kaa02}; $^{f}$\citet{tru02}; $^{g}$\citet{bok03}; $^{h}$\citet{bko03}; $^{i}$\citet{man04}; $^{j}$W04; $^{k}$\citet{bko04}; $^{l}$\citet{tru04}

\end{table*}

Those PDS that were well fitted by broken power law models, when power law models could be rejected at a level $\ga$99.9\%, were classified as Type A.  We also classified as Type A those PDS where the probability of improved fitting with a broken power law was  $>$99.9\%. PDS that were well described by a simple power law, and clearly displayed power in excess of the Poisson noise, were classified as Type B. Those PDS which exhibited no excess power were classed as flat.  Any PDS that were not classifiable by any of the above means were rejected. We only accepted PDS classifications where the 64 bin and 128 bin PDS  of the same lightcurve were classified as the same type (i.e. both Type A, both Type B or both flat).

We simulated random lightcurves with Type B  PDS following the method of \citet{tk95} and normalised the intensities to 0.5 count s$^{-1}$  (a typical  combined EPIC intensity of a bright X-ray binary in M31). We found that the power often dipped below the Poisson noise level in the observed frequency range. Hence, a flat PDS with an  average source intensity that was higher than the faintest  Type A source in the same observation  was consistent with Type B but not Type A.  

\begin{figure}[!t]
\resizebox{\hsize}{!}{\includegraphics[angle=270]{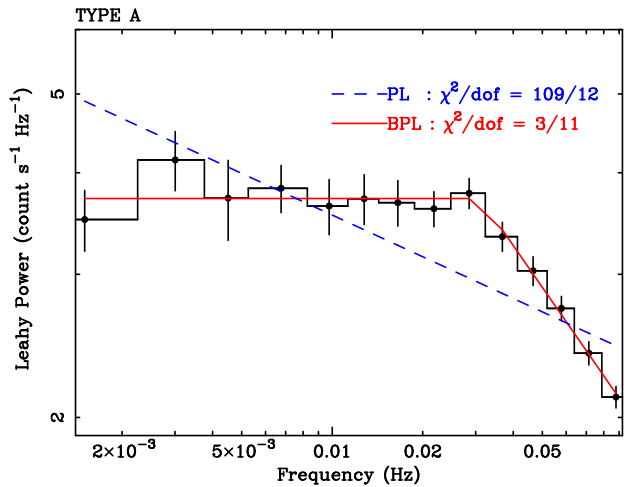}}
\caption{Example Type A PDS, exhibited by { S45} in Observation 4; the PDS is averaged over 96 intervals of 128 bins, and grouped. The axes are log-scaled. The PDS is  Leahy normalised, so that the Poisson noise has a power of 2. Fits of power law (PL) and broken power law (BPL) models to the PDS are shown; the power law fit is clearly unacceptable, while the broken power law fits well.}\label{ta}
\end{figure}

\begin{figure}[!t]
\resizebox{\hsize}{!}{\includegraphics[angle=270]{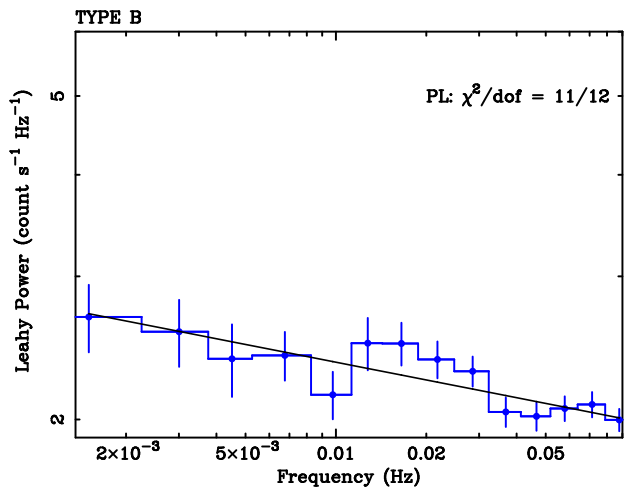}}
\caption{Example Type B PDS exhibited by S32 in Observation 4; the PDS is averaged over 96 intervals of 128 bins, and grouped. The axes are log-scaled. The PDS  is Leahy normalised, so that the Poisson noise has a power of 2.}\label{tb}
\end{figure}

Source and background SEDs  were extracted for each EPIC-pn observation of every source in 4096 channels, with 5 eV binning. A response matrix was generated for each source, and corresponding ancillary response files were also made. Counts outside the 0.3--10 keV range were rejected, and the spectra were grouped  according to the source intensity: spectra exceeding 500 counts were grouped for a minimum of 50 counts per bin, and spectra containing less than 500 counts were grouped to a minimum of 20 counts per bin.

\section{Results}
\label{varsum}
The 63 sources that satisfied the intensity selection in at least one of the four observations are catalogued in Table~\ref{prop}. The J2000 coordinates are given, along with their designations in the first and second ROSAT PSPC catalogues (S97, S01) and the Chandra catalogues of K02 and \citet[][]{will04}, hereafter referred to as  W04; known properties of each source are given, with separate columns for the observed PDS types and the most likely  primaries where known.  The coordinates are taken from K02, or W04 for sources not in the K02 catalogue. Intensity  variability over short ($\sim$100 s) timescales is indicated by $v$, and X-ray sources that vary in luminosity by more than 5$\sigma$ between observations  are labelled $t$. We define transients (labelled $T$) as sources that vary in intensity by a factor of $\ga$10, and are absent in at least one observation. Foreground sources are indicated by $fg$, globular cluster sources by $glob$, supernova remnants by $snr$ and supersoft sources by $sss$. Dipping is indicated by $dip$, Z-source branch movement by $zs$, pulsations by $pulse$, and positional coincidence with planetary nebulae by $pn$. A likely neutron star primary,  as prescribed in Table~\ref{class},  is indicated by $ns$, while  a black hole candidate is indicated by $bh$. Radio counterparts are indicated by $r$, and are taken from the catalogues of \citet{walt85} and \citet{walt02}.    K02 found positional coincidences with the globular clusters using the catalogues of \citet{bat87},  \citet{mag93}, and \citet{barm00}. The planetary nebulae were assigned using the catalogues of \citet{fj78}, and \citet{cjfn89} --- K02 comment that these X-ray sources  are 3 orders of magnitude brighter than for PN.
 \citet{wgm04} obtained the X-ray positions of the six known PN associations from a 37 ks Chandra ACIS observation; they found that the planetary nebulae are not responsible for the X-rays, but are likely to be associated with the X-ray emitter. The reasons for these associations are not clear.
 The SNR were identified from \citet{ddb80}, \citet{bw93}, and \citet{mag95}. Point sources in XMM-Newton that are resolved by Chandra into multiple sources are designated $m(x)$, where $x$ is the number of sources observed by Chandra. Finally, the types of variability are indicated: $A$ for Type A, $B$ for Type B,  and $-$ for flat PDS. If more than one type of variability is observed from a given source, then they are observed in different observations.

In the following sections, we first describe the properties of PDS observed from the population as a whole (Sect.~\ref{respds}). We present luminosity functions of Type A, Type B and flat PDS, and use them to identify possible black hole candidates. We then discuss new insights into the globular cluster X-ray sources in Sect.~\ref{globclust}, supernova remnants in Sect.~\ref{snr}, our black hole candidates in Sect.~\ref{bhc}, and supersoft sources in Sect.~\ref{sss}. Our results shed no new light on the known dipping sources.

\subsection{Power density spectra}
\label{respds}

We provide a  description of the PDS and SEDs for each observation of every source in Appendix~\ref{appendix}.  We were able to classify 107 of the 252 power density spectra.
  Examples of Type A and Type B PDS are provided in Fig.~\ref{ta} and Fig.~\ref{tb} respectively. The axes of these figures are logarithmic, and the power is Leahy normalised so that the Poisson noise has a power value of 2. The power is measured in units of count s$^{-1}$ Hz$^{-1}$.

 In Fig.~\ref{ta} we present a 128 bin  PDS of { S45} from Observation 4, averaged over 96 intervals. Best fit power law (dashed line) and broken power law (solid line) models are presented, with $\chi^2$/dof = 109/12 and 3/11 respectively.
Figure~\ref{tb} shows  the equivalent PDS for S32 in Observation 4. The PDS is well described by a simple power law, with $\chi^2$/dof = 11/12. The { spectral} index, $\gamma$ = $-$0.07$\pm$0.01,  is shallower than expected for Galactic LMXBs in a Type B state; however, it is consistent with the PDS of our simulated Type B lightcurves.

\begin{figure}[!t]
\resizebox{\hsize}{!}{\includegraphics{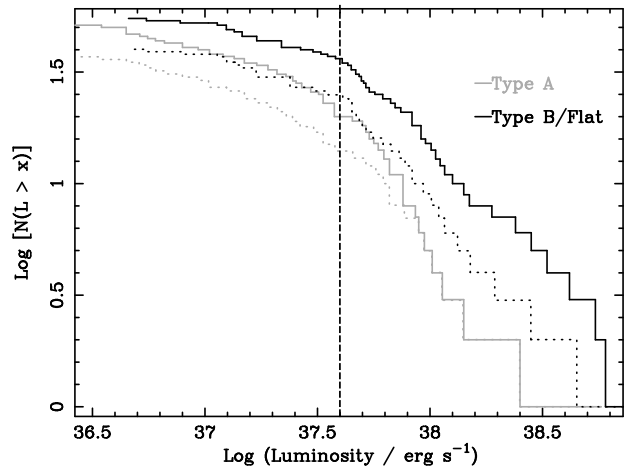}}
\caption{ Cumulative  luminosity functions for Type A (grey) and Type B/flat PDS (black) in the 0.3--10 keV band. { Dotted  lines indicate luminosity functions where each source is only counted once, while solid lines indicate that all the data are used}. The dotted vertical line indicates  a luminosity of 4$\times$10$^{37}$ erg s$^{-1}$, the theoretical maximum luminosity for Type A variability  in a neutron star LMXB(see text). Indeed, we see that the LF above  4$\times$10$^{37}$ erg s$^{-1}$ is considerably steeper than below. The shallow and steep parts of the LF could represent the neutron star and black hole populations, respectively.}\label{cdfa1}
\end{figure}

\begin{figure}[!t]
\resizebox{\hsize}{!}{\includegraphics{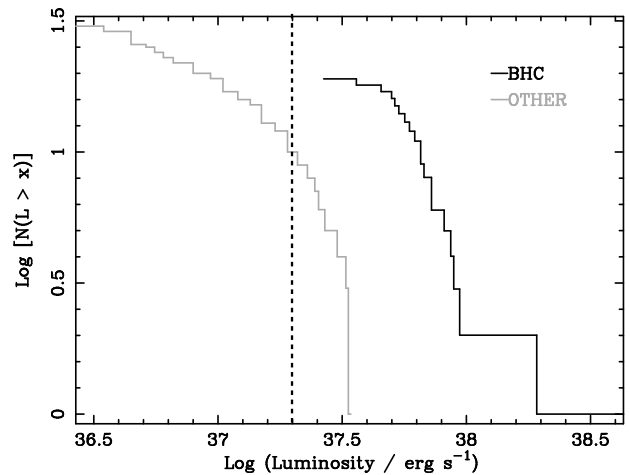}}
\caption{  Cumulative luminosity functions for Type A PDS in black hole candidates (black), and other sources (grey). We classify as black hole candidates those X-ray sources that are resolved by Chandra as a single point, and exhibit Type A PDS at a 0.3--10 keV luminosity $>$ 4$\times$10$^{37}$ erg s$^{-1}$ in at least one observation. { The luminosity $L_{\rm c}$ = 0.1 $L_{\rm Edd}$ for 1 4. M$_{\odot}$ neutron star is indicated by a vertical, dashed line.}}\label{cdfa2}
\end{figure}

\subsubsection{Type A PDS}
\label{typea}
We observed  53 Type A PDS in total, over a luminosity range of $\sim$5--400$\times$10$^{36}$ erg s$^{-1}$ in the 0.3--10 keV band.
 The break frequencies of our Type A PDS range over $\sim$0.02--0.06 Hz,
which is in the range 0.01--1 Hz found in Galactic LMXBs by \citet{vdk94}.  However, we note that these break frequencies are systematically lower than those observed in Galactic LMXBs, which are typically $\sim$0.1 Hz and usually in the range 0.04--0.4 Hz \citep{vdk95}. This may suggest that the M31 X-ray population may differ systematically from the Galactic LMXB population, although we do not believe this to be the case necessarily. The observed range of $\beta$ was $\sim$0.2--1.9; however $\sim$80\% were consistent with $\beta$ $\sim$0.5--1. Hence, broadly speaking,
$\beta$ $\sim$1, consistent with Galactic LMXBs. However, our observed
values of $\beta$ are systematically lower than in the Galactic LMXBs;
this is likely to be due to the low count rate. 

 \begin{table}[!t]   
\renewcommand{\baselinestretch}{1}
\renewcommand{\tabcolsep}{1mm}
\caption{\footnotesize   Fits to the luminosity functions presented in Figs.~\ref{cdfa1}--\ref{cdfb}. Column 1 gives the LF to be fitted, and the model used: power law (PL) or broken power law (BPL).   Column 2 gives the relevant figure number. $\alpha$  and $\beta$ are the two spectral indices below and above the break luminosity $L_{\rm b}$. Prob. is  the probability that the model is a good fit to the LF. Numbers in parentheses are 90\% confidence level uncertainties on the last digit.}\label{lffits}
\begin{tabular}{lllllllllllll}
\noalign{\smallskip}
\hline
\noalign{\smallskip}
LF$^a$ &  Fig. & Model & $\alpha$  & $\beta$& $L_{\rm b}$$^b$ & Prob.$^c$\\
\noalign{\smallskip}
\hline
\noalign{\smallskip}
A Tot & 3&   PL& 0.52(8) & $\dots$ & $\dots$&  0.01\\
A Tot && BPL  & 0.34(9) & 1.8(5) & 4.4(7) & 0.81\\
A BH &    4 & PL & 0.40(10) & $\dots$ & $\dots$ &  2$\times$10$^{-4}$ \\
A BH & &BPL & 0.05(12) & 1.5(5) & 4.6(7) &  0.97\\
A NC &  4  & PL & 0.8(2) & $\dots$ & $\dots$ & 0.07\\\vspace{0.07in}
A NC & &BPL & 0.5(2) & 3.6(19) & 2.1(4) & 0.89\\

B Tot & 3& PL & 0.55(2)&  $\dots$ & $\dots$ &  0.09\\
B Tot & &BPL & 0.30(2) & 0.97(7) & 3.6(3) & 0.68\\
B BH & 6 &  PL & 1.7(11) &$\dots$ & $\dots$& 0.93\\
B BH & &BPL & 2.8(2) & 1.9(8) & 4.3(17) & 0.52\\
B NC & 6 & PO & 0.53(7) & $\dots$ & $\dots$& 0.09\\
B NC &&  BPL & 0.36(10) & 0.8(2) & 5(2) & 0.56\\

\noalign{\smallskip}
\hline
\noalign{\smallskip}
\end{tabular}
\\{\footnotesize $^a$ Type A (A) and Type B/flat (B) LFs were modeled for the total (Tot), black hole (BH) and non-classified (NC) populations. }

{\footnotesize $^b$ 0.3--10 keV luminosity normalised to 10$^{37}$ erg s$^{-1}$} 

{\footnotesize $^c$ The goodness of fit was obtained using the method outlined in \citep{cjm70}}

\end{table}

{ We present  0.3--10 keV luminosity functions (LFs) for  the Type A PDS (in grey) in Fig.~\ref{cdfa1}.  In the solid LF several sources are represented more than once, since they exhibit  Type A variability in more than one observation; in the dotted LF each source is represented only once, by its highest luminosity Type A variability.
 The dashed vertical line indicates a luminosity of 4$\times$10$^{37}$  erg s$^{-1}$, which we consider to be the maximum luminosity for Type A variability in a neutron star (Sect.~\ref{litreview}). We classify as a black hole candidate any source that exhibits Type A variability at a 0.3--10 keV luminosity $>$4$\times$10$^{37}$  erg s$^{-1}$ in at least one observation. The slope of the LF is considerably shallower below  4$\times$10$^{37}$  erg s$^{-1}$ than above, which may indicate two populations, neutron star Type A's at low luminosity and black hole Type A's at high luminosity. 

{  We modelled the Type A LF with two models: a power law and a broken power law. We ignored the Type A variability of S56, as this appears to be a supersoft source (See Sect.~\ref{sss}). The fitting was performed with the {\sc Sherpa} modelling package, using Cash statistics \citep{cash79}. We obtained the goodness of fit using the method outlined in \citet*{cjm70} for unbinned data without uncertainties. 

We present the results of fitting the LF in Table~\ref{lffits}. Uncertainties are quoted at the 90\% confidence levels. We find that the Type A LF cannot be  described by a single power law; the best fit  has a 1\% probability of successfully describing the data. However, it is well described by a broken power law where the spectral index changes from 0.34$\pm$0.09 to 1.8$\pm$0.5 as the luminosity exceeds 4.4$\pm$0.7$\times$10$^{37}$ erg s$^{-1}$ in the 0.3--10 keV band. This break may indicate the boundary between neutron star and black hole populations, lending support to our classification of sources  as black hole LMXBs if they exhibit Type A variability at 0.3--10 keV luminosities $>$ 4$\times$10$^{37}$ erg s$^{-1}$. 
}

In determining our observed luminosities, we assume that the emission is isotropic. Instead, the emission could be anisotropic and beamed in our direction. In this case, the range of high luminosity Type A variability could be due to varying beaming factors rather than variation in primary masses. This possibility is further explored in Sect.~\ref{beaming}.

In Fig.~\ref{cdfa2}, we separate the Type A LF into black hole candidates and  other X-ray sources, most likely neutron star LMXBs. However, this non-classified population may contain hidden black holes where $l$ $\ll$ $l_{\rm c}$. Here, the dashed line represents 2$\times$10$^{37}$ erg s$^{-1}$, which is our value of $L_{\rm c}$ for a 1.4 M$_{\odot}$ neutron star.  { Fits to the black hole and non-classified Type A LFs are presented in Table~\ref{lffits}.

The Type A LF for the black hole candidate population appears well described by a simple power law except at the low luminosity end where we are necessarily incomplete.  However, a  power law fit is rejected because some of the black hole candidates also exhibited Type A variability at luminosities $<$4$\times$10$^{37}$ erg s$^{-1}$. Instead, the LF is well described by a broken power law with the spectral index changing from 0.05$\pm$0.12 to 1.5$\pm$0.5 at a luminosity of 4.6$\pm$0.7$\times$10$^{37}$ erg s$^{-1}$.

The Type A LF for the non-classified population is best described by a broken power law with the spectral index changing from 0.5$\pm$0.2 to 3.6$\pm$1.9 at a luminosity of 2.1$\pm$0.4$\times$10$^{37}$ erg s$^{-1}$. Interestingly, this break agrees well with our predicted maximum luminosity for Type A variability in a 1.4 M$_{\odot}$ neutron star, suggesting that this non-classified population is indeed dominated by neutron star LMXBs. However, the best fit power law model has a goodness of fit probability of 7\%, and hence cannot be rejected. 
}

We also note that $\sim$45\% of the  unidentified Type A PDS were observed at less than 1.0 $\times$ 10$^{37}$ erg s$^{-1}$; this strongly  argues against the presence of a bias towards higher luminosity (i.e. signal to noise) that would occur if the Type A phenomenon was due to statistical rather than intrinsic noise in the systems.
\begin{figure}[!b]
\resizebox{\hsize}{!}{\includegraphics[angle=270]{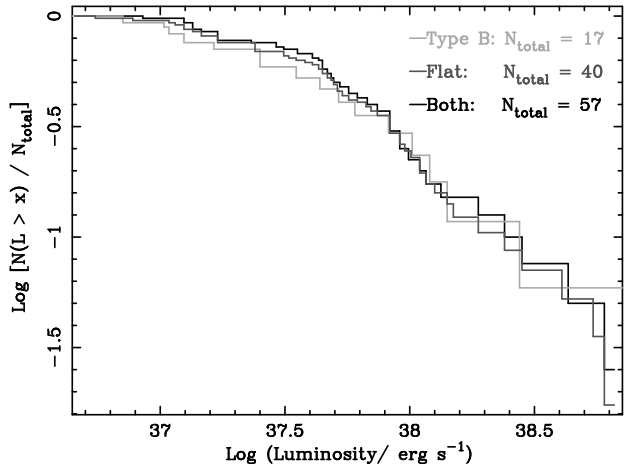}}
\caption{Normalised, fractional luminosity functions for Type B (light grey) and  flat (dark grey) PDS.  KS testing shows that these two populations have a 47\% chance of being drawn from the same distribution. The similarity between these supports the idea that the flat PDS are consistent with Type B PDS.  The combined fractional luminosity function (black) of Type B and flat PDS  is also given.}\label{lbf}
\end{figure}

\begin{figure}[!b]
\resizebox{\hsize}{!}{\includegraphics[angle=270]{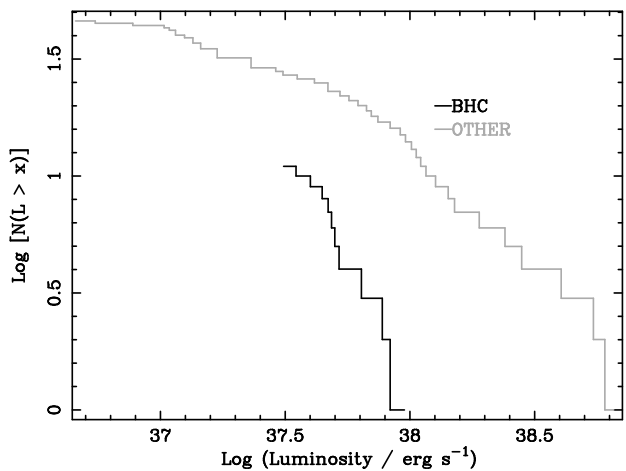}}
\caption{  Cumulative luminosity functions for Type B/flat  PDS in black hole candidates (black), and other sources (grey). The axes are log-scaled. The lowest luminosity Type B/flat PDS for the black hole candidates is consistent with $\sim$4$\times$10$^{37}$ erg s$^{-1}$ (in S63, Table~\ref{allprops}). This result supports $l_{\rm c}$ $\sim$0.1.  We note that there are no observations of Type B variability in the black hole candidates at 0.3--10 keV luminosities $>$10$^{38}$ erg s$^{-1}$. This is not unexpected, since Galactic black hole LMXBs are observed in the steep power law state at high luminosities, and exhibit Type A variability.
}\label{cdfb}
\end{figure}

\subsubsection{ Type B / Flat PDS}

In Fig.~\ref{lbf}, we present luminosity functions of  Type B and flat PDS, normalised by the total number in each sample.  KS-testing gave a probability of 47\% that the two LFs were drawn from the same distribution. Hence, we consider  the  Type B and flat PDS to be consistent with being from the same distribution, suggesting that most of the flat PDS in our sample in fact come from high accretion rate LMXBs.

The combined  luminosity function of flat and Type B PDS is also shown in Fig.~\ref{lbf}.   This total LF for Type B and flat PDS is compared with the Type A LF in Fig.~\ref{cdfa1}. { Best fit power law and broken power law models for the total Type B/flat LF are presented in Table~\ref{lffits}.  The best fit power law model provides an acceptable fit (9\% probability of a good fit); however, a broken power law is prefered (68\% probability of a good fit). The spectral index of the best fit broken power law model changes from 0.30$\pm$0.02 to 0.97$\pm$0.07 at 3.6$\pm$0.3$\times$10$^{37}$ erg s$^{-1}$.  We see from Fig.~\ref{cdfa1} and Table~\ref{lffits} that the Type A and Type B/flat LFs are distinct. Hence we can be confident that the break in the Type A LF is not some artifact of the XMM-Newton observations.}

Figure~\ref{cdfb} shows the combined Type B/flat  luminosity functions} for the  black hole candidate and unidentified  populations. Even though the black hole candidates were selected for their Type A PDS,  no Type B/flat PDS is observed below $\sim$4$\times$10$^{37}$ erg s$^{-1}$ in this  black hole population. Hence these results support $l_{\rm c}$ $\sim$0.1. { We see from Table~\ref{lffits} that the Type B/flat LF for the black hole candidates is best described by a power law with spectral index 1.7$\pm$ 1.1. The Type B/flat LF for the non-classified population is consistent with that of the total population. 
}

 In Fig.~\ref{cdfb} we see four observations of Type B/flat PDS in the  unidentified population at 0.3--10 keV luminosities in excess of 3$\times$10$^{38}$ erg s$^{-1}$, well above L$_{\rm Edd}$ for a 1.4 M$_{\odot}$ neutron star. Three of these are observations of the Z-source S26; however, the emission is locally sub-Eddington, with the neutron star contributing only 7--15$\times$10$^{37}$ erg s$^{-1}$ \citep{bko03}. The remaining emission was contributed by the ADC, which is probably extended (see Sect. 2), and hence locally sub-Eddington also.  The other high luminosity observation was from the supersoft source S35, which may be beamed (see Sect. 7.5).

\subsubsection{Variability of the PDS}
\label{fvar}

The PDS of Galactic LMXBs are characterised by their r.m.s. variability, as detailed in Sect.~\ref{sedpds}. We classified our M31 X-ray sources using the fractional r.m.s. variability amplitude, $F_{\rm var}$,   accounting for the uncertainties in the measurements \citep[see e.g.][]{vgn03}. We then compared the  variability of Type A and Type B/flat PDS in our sample with the known r.m.s variability of Galactic LMXBs.

The r.m.s. variability of our sample is measured in the 0.001--0.1 Hz range, while r.m.s. variability of Galactic LMXBs is often measured in the 0.1--10 Hz range \citep[see e.g.][]{mr03}.
The PDS and r.m.s. variability properties of Galactic LMXBs are considerably  better known than those of our sample; hence, we estimated the r.m.s. variabilities of Galactic LMXBs in the 0.001--0.1 Hz range. The r.m.s. variability of a lightcurve is related to the integrated PDS \citep[see e.g.][]{vgn03}. Hence, if $V_{\rm lo}$ and $V_{\rm hi}$ represent  the source variability in the 0.001--0.1 Hz and 0.1--10 Hz ranges respectively, and $A_{\rm lo}$ and $A_{\rm hi}$ represent the area under the PDS in the same frequency ranges, then 
\begin{equation}
\frac{V_{\rm lo}}{V_{\rm hi}} \sim \frac{A_{\rm lo}}{A_{\rm hi}}.
\end{equation}

To estimate $V_{\rm lo}$/$V_{\rm hi}$, we created artificial Type A PDS with the spectral index $\gamma$ changing from 0 to 1 at  break frequencies of 0.01 and 0.07 Hz, spanning the frequency range of observed break frequencies within 90\% confidence limits (see Table A.2). We found $V_{\rm lo}$/$V_{\rm hi}$ to be $\sim$0.7 and 0.3 respectively. This suggests that  we are most sensitive to systems with low break frequencies. We also created Type B PDS, using a simple power law model with { $\gamma$ =  1.0 and 1.5, spanning the range of spectral indices for Type B variability in Galactic LMXBs,   giving $V_{\rm lo}$/$V_{\rm hi}$ $\sim$1.1 and $\sim$13 respectively.}
 However, we will need to examine the 0.001-0.1 Hz PDS of Galactic LMXBs in detail at a later date. 

\begin{figure}[!t]
\resizebox{\hsize}{!}{\includegraphics[angle=270]{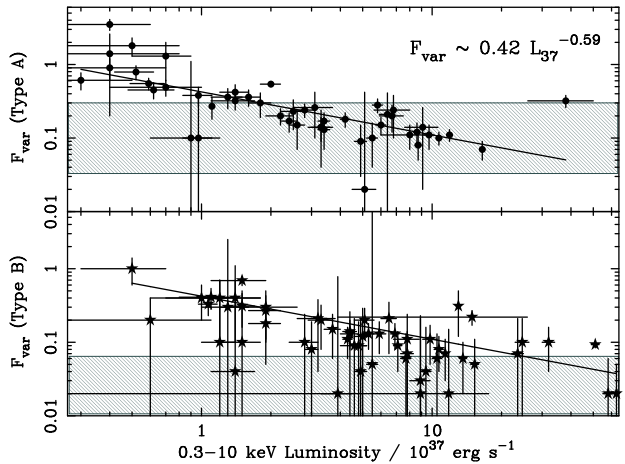}}
\caption{ Fractional r.m.s. variability, $F_{\rm var}$ vs. 0.3--10 keV luminosity for Type A (top panel) and Type B/flat (bottom panel) PDS. The axes are log-scaled. Shaded regions show estimated r.m.s. variabilities for Galactic Type A and Type B PDS in the 0.001-0.1 Hz range. A power law fit to the Type A data is shown in both panels, for comparison of the variabilities of Type A and Type B/flat PDS. }\label{rmsl}
\end{figure}

We present plots of  $F_{\rm var}$ vs. 0.3--10 keV luminosity for the classified PDS in Fig.~\ref{rmsl}; the results for Type A PDS are shown in the top panel, and Type B/flat results are shown in the bottom panel. The error bars represent the 90\% confidence limits. In each panel,  the shaded area shows the range of r.m.s. variability for Galactic LMXBs in the 0.001-0.1 Hz range, estimated as  above. We also plot in each panel the best fit power law model to the Type A data.

 \begin{table*}[t]   
\renewcommand{\baselinestretch}{1}
\caption{\footnotesize  Luminosities, SED shapes and PDS for each of the 14 bright globular cluster X-ray sources in Observations 1--4. $L_N$ is the 0.3--10 keV luminosity in Observation $N$, in units of 10$^{37}$ erg s$^{-1}$; $S_{N}$ is the photon index for the best to the SED of a power law model, and $T_N$ is the PDS type. ``$\dots$'' indicates that no good SED was obtained; either the source was not observed with the PN, or the SED had negative counts.}\label{globstats}
\begin{tabular}{lllllllllllll}
\noalign{\smallskip}
\hline
\noalign{\smallskip}
{ Source }&{ $L_1$ } & $S_1$ &{ $T_1$ } & $L_2$ & $S_2$ & $T_2$ & $L_3$ & $S_3$ &  $T_3$ & $L_4$ &$S_4$ &{ $T_4$}\\
\noalign{\smallskip}
\hline
\noalign{\smallskip}
S4 & 1.3(3) & 2.2(4) & $-$ & 1.0(3) & 1.7(5) & B & 1.2(3) & 2.2(3) & ? & 1.8(2) & 2.08(16) & ?\\
S5 & 0.9(5) & 2.7(4) & ? & 1.0(3) & 2.0(4) & ? & 1.2(2) & 2.0(4) & ? & 3.6(3) & 1.91(13) & ?\\
S9 & 24.6(17) & 1.39(9) & $-$ & $\dots$ & $\dots$ & $\dots$& 23.5(14) & 1.48(9) & $-$ & $\dots$ & $\dots$ & $\dots$\\
S11 & 9.4(7) & 1.40(9) & ? & 7.3(9) & 1.33(13) & ? & 4.4 & 1.56(8) & $-$ & 7.0(3) & 1.59(6) & ?\\
S15 & 1.9(3) & 2.0(3) & B & 1.2(3) & 2.4(5) & ? & 1.9(3) & 2.3(2) & ? & 1.4(2) & 2.3(2) & ? \\
S21 & 1.0(2) & 1.8(2) & ? & 3.2(6) & 2.6(4) & $-$ & 3.4(3) & 2.0(13)&  ? & 2.2(3) & 2.15(9) & A \\
S42 & 4.4(3) & 2.18(14) & ? & 5.0(7) & 2.02(13) & ? & 4.9(3) & 1.87(9) & $-$ & 5.2(2) & 2.02(7) & ?\\
S43 & 4.8(4) & 1.56(10) & $-$ & 5.1(6) & 1.7(2) & A & 5.5(3) & 1.58(8) & A & 5.8(2) & 1.58(6) & ?\\
S44 &  $\dots$ & $\dots$ & $\dots$ &$\dots$ & $\dots$ & $\dots$ & $\dots$ & $\dots$ & $\dots$ & 6.5(5) & 0.98(10) & $-$\\
S47$^{a}$ & 3.5(3) & 2.28(15) & ? & 4.6(6) & 1.9(2) & $-$ & 3.4(2) & 2.01(11) & A & 11.9(3) & 1.89(4) & A\\
S48 & 1.9(3) & 2.2(2) & ? & 1.1(3) & 2.1(3) & ? & 1.6(2) & 2.3(3) & ? & 1.6(2) & 2.2(2) & A \\
S49 & $\dots$ & $\dots$ & ? & 3.9(6) & 1.9(3) & ? &   $\dots$ & $\dots$ & ? & 2.4(2) & 1.87(13) & A \\
S54 & 13.3(6) & 1.58(6) & ? & 13.2(9) & 1.71(11) & ? & 11.4(4) & 1.79(4) & $-$ & 10.3(3) & 1.72(2) & ? \\
S55$^b$ & 14.9(11) & 0.64(5) & B & 21(6) & 2.1(9) & ? & 22(3) & 2.4(3) & ? & 12.1(8) & 0.52(7) & ?\\
S62 & 8.4(7) & 1.42(10) & ? & 6.6(8) & 1.5(2) & ? & 7.7(5) & 1.62(10) & ?  & 6.8(3) & 1.70(7) & ?\\

\noalign{\smallskip}
\hline
\noalign{\smallskip}
\end{tabular}
\\$^a$ S47 is resolved into two sources by Chandra, the globular cluster X-ray source to the south, and a black hole candidate to the north.\\
$^b$ The spectra  of S55 in observations 1 and 4 are particularly hard because non-dip and dip spectra are mixed.\\\end{table*}

We see that the variabilities of most of the PDS are consistent with their Galactic counterparts.  The Type B PDS are systematically more variable than Galactic counterparts with $\gamma$ = 1.0; however, they are consistent with the known range of $\gamma$ for Galactic Type B variability:  1.0 $\le$ $\gamma$ $\le$ 1.5 \citep{vdk94}. The Type A variability is broadly characterised by $F_{\rm var}$ $\sim$0.42 $L_{37}^{-0.59}$, where $L_{37}$ is the 0.3--10 keV luminosity in units of 10$^{37}$ erg s$^{-1}$. We see from the bottom panel of Fig.~\ref{rmsl} that 
$F_{\rm var}$ $<$0.42 $L_{37}^{-0.59}$ for most Type B PDS; hence our Type A PDS tend to be  more variable than our Type B PDS as expected, but this result is not strongly significant.

\subsection{The bright globular cluster X-ray sources}
\label{globclust}
 There are thirteen bright X-ray sources that have been identified in Galactic globular clusters. Twelve of these  have been identified as neutron star LMXBs, while the thirteenth has not been classified \citep[][ and references therein]{int04}. Hence the 14 globular cluster X-ray sources in our sample are expected to be LMXBs containing $\sim$1.4 M$_{\odot}$ neutron stars.
This makes their observed spectral and timing behaviour of particular interest.

We summarise the spectral and timing properties of the globular cluster sources in Table~\ref{globstats}. We stress that S47 is resolved into two unrelated sources by Chandra; the southern source is associated with the globular cluster, while the northern source is transient and contains a black hole candidate \citep{bko04}.  { Our results for the globular cluster sources are consistent with Type A variability being observed at lower luminosities than Type B.  Furthermore, our SEDs are well described by power laws with photon index $\sim$1.5--2.1 when Type A variability is observed,  consistent with LMXBs in the low/hard state as classified by \citet{mr03}. In this respect, the globular cluster X-ray sources behave in a way consistent with Galactic neutron star LMXBs.}

 In Fig.~\ref{globfig} we show PDS type vs. luminosity for observations of the globular cluster sources where the PDS is classified; S11 is omitted because it showed no classified variability.  As before,  we assume that $l_{\rm c}$ $\sim$0.1 in the 0.3--10 keV band.  A dashed vertical line represents a luminosity of 2$\times$10$^{37}$ erg s$^{-1}$, i.e. $\sim$10\% Eddington for a 1.4 M$_{\odot}$ neutron star. Out of the 12 globular cluster X-ray sources in Fig.~\ref{globfig},  3 systems show more than one type of variability (S4, S21, and S43), { while 6 systems are observed at luminosities near } $L_{\rm c}$ $\sim$2$\times$10$^{37}$ erg s$^{-1}$ (S4, S15, S21,  S43, S48, S49).  Hence we can only estimate $l_{\rm c}$ with these data. { Ten of the twelve sources behaved as expected for LMXBs with a 1.4 M$_{\odot}$ neutron star primary. However, S4 and S43 appear at face value to be discrepant. We discuss how these sources may fit into our scheme below.
 
\begin{figure}[!b]
\resizebox{\hsize}{!}{\includegraphics[angle=00]{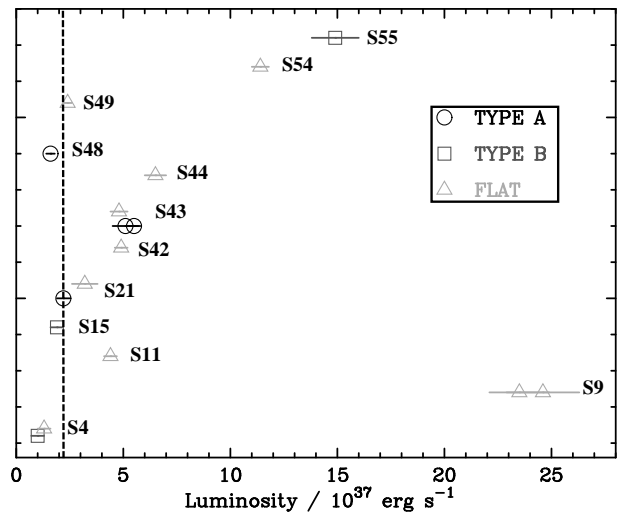}}
\caption{  PDS type and 0.3--10 keV luminosity for observations of globular cluster X-ray sources where the PDS type is classified. A dashed line represents 2$\times$10$^{37}$ erg s$^{-1}$,  i.e. 10\% Eddington for the assumed 1.4 M$_{\odot}$ primary.  The y-axis is used to separate the results of the different sources and has no physical meaning. Circles, squares and triangles represent observations of Type A, Type B and flat PDS respectively. We expect Type A PDS at luminosities below,  and Type B or flat PDS above, 2$\times$10$^{37}$ erg s$^{-1}$. {  S43 is inconsistent with $L_{\rm c}$ = 2$\times$10$^{37}$ erg s$^{-1}$, as it exhibited Type A variability at a 0.3--10 keV luminosity of 5--6$\times$10$^{37}$ erg s$^{-1}$. S4 exhibited Type B and flat PDS at lower luminosities than expected; this may be due to a low mass primary, or hysteresis.}}\label{globfig}
\end{figure}

Most striking are the two observations of Type A variability in S43, at luminosities $>$5$\times$10$^{37}$ erg s$^{-1}$. A third observation of S43 has a flat PDS, at a luminosity that is consistent within errors with being greater than the Type A luminosity in S43. Hence, S43 is consistent with $l_{\rm c}$ $\sim$0.1 if it contains a black hole rather than a neutron star.
 Although no Galactic black hole LMXBs { have been identified} in globular clusters, we  note that \citet{ang01} have reported a possible globular cluster black hole binary in the elliptical galaxy \object{NGC\thinspace 1399}, with a 0.3--10 keV luminosity of 5$\times$10$^{39}$ erg s$^{-1}$.

 Other  possibilities are explored by \citet{dis02} when discussing an M31 globular cluster X-ray source { in the disc} with a 0.2--7 keV luminosity of  2--6$\times$10$^{38}$ erg s$^{-1}$. Firstly, the globular cluster could contain two or more X-ray sources. \citet{wa01} discovered one globular cluster with a second bright X-ray source out of a population of 13 Galactic X-ray bright globular clusters, hence it is reasonable to expect that $\sim$1 of the 14 X-ray bright globular clusters in our sample contains multiple LMXBs. { Alternatively, S43 could contain a 1.4 M$_{\odot}$ neutron star primary, accreting $\sim$100\% helium, rather than hydrogen; this would raise the Eddington limit by a factor of 2. } Finally, { the observed luminosity could be artificially enhanced by relativistic beaming, or overestimated if anisotropic}; this will be discussed in Sect.~\ref{beaming}. 
 
{  S4 exhibited flat and Type B PDS  in observations 1 and 2 respectively, at 0.3--10 keV luminosities of 1.3$\pm$0.3 and 1.0$\pm$0.3 $\times$10$^{37}$ erg s$^{-1}$. According to our model, Type A PDS are expected at these luminosities. S4 is consistent with $l_{\rm c}$ = 0.07 $L_{\rm Edd}$ for a 1.4 M$_{\odot}$ neutron star primary. However, S4 would be also consistent with $l_{\rm c}$ = 0.1 for a LMXB with a $\sim$1.0 M$_{\odot}$ neutron star primary. There is already observational evidence for systems with low mass neutron stars. An example is the eclipsing binary pulsar SMC X-1, which is discussed next.

\begin{table*}[!t]   
\renewcommand{\baselinestretch}{1}
\renewcommand{\tabcolsep}{1mm}
\caption{ Luminosities, SED shapes and PDS  in Observations 1--4  for each of the 14 black hole candidates, either exhibiting Type A variability at 0.3--10 keV luminosities $>4\times 10^{37}$ erg s$-1$ in any observation, or identified by some other method. $L_N$ is the 0.3--10 keV luminosity in Observation $N$, in units of 10$^{37}$ erg s$^{-1}$; $S_{N}$ is the photon index for the best to the SED of a power law model, and $T_N$ is the PDS type. ``$\dots$'' indicates that no good SED was obtained; either the source was not observed with the PN, or the SED had negative counts.}\label{lspec}
\begin{tabular}{lllllllllllll}
\noalign{\smallskip}
\hline
\noalign{\smallskip}
 Source & $L_1$ &$S_1$ & $T_1$ & $L_2$ & $S_2$&$T_2$ & $L_3$ & $S_3$ & $T_3$ & $L_4$ &$S_4$ & $T_4$\\
\noalign{\smallskip\hrule\smallskip}
S7 & 3.7(3) & 2.3(2) & B & 4.9(5) & 2.0(3) & ? & 5.3(3) & 2.09(10) & B & 8(2) & 2.47(12) & A \\
S12 & 4.8(5) & 2.21(15) & ? & 5.0(6) & 2.2(3) & ? & 4.6(3) & 2.14(13) & ? & 6.0(3) & 2.18(9) &A\\
S14 & 7.7(4) & 1.73& $-$$^a$ & 5.7(7) & 1.7(2) & ? & 24.6(12) & 2.39(7) & ?$^a$& 23.8(5) & 2.33(6) & ?$^{a}$\\ 
S19 & 3.3(4) & 2.9(4) & B & 2.9(6) & 2.3(2) & ? & 5.1(5) & 3.0(3) & B & 9.7(5) & 2.47(9) & A\\
S37 & 9.3(6) & 1.88(8) & ? & 8.9(7) & 1.84(12) & $-$ & 8.6(3) & 1.89(6) & A & 8.7(3) & 1.84(5) & A\\
S40 & 4.2(3) & 2.02(13) & ? & 4.3(5) & 2.1(3) & ? & 4.9(3) & 1.88(9) & A & 4.3(2) & 2.09(8) & $-$\\
S43$^{b}$ & 4.8(4) & 1.56(10) & $-$ & 5.1(6) & 1.7(2) & A & 5.5(3) & 1.58(8) & A & 5.8(2) & 1.58(6) ?\\
S47$^{c}$ & $\sim$0 & ? & ? & $\sim$2 &? & ? & $\sim$1.7 & ? & A & $\sim$5.3 & ? & A\\
S51$^{d}$ & 0.08(8) & 5(3)$^{e}$ & ? & 0.8(8) & 3(6)$^{e}$  & ? & 0.4(4) & 4(4)$^{e}$  & A & 5.5(3) & 2.30(11) & ?\\
S53 & 0.21(14) & 3.5(8) & ? & 0.2(2) & 3.5(8) & ? & 0.16(12) & 3.6(13) & ?& 4.2(2) & 1.83(8) & A\\
S56  & 36(15) & 9$^{e}$  & ?& 3(2) &  4$^{e}$  & ? & 1.5(10) & 5(2)$^{e}$  & ? & 38(12) & $\sim$8$^{e}$  & A\\
S59 & 7.8(10) & 0.86(16) & $-$ & 9.1(15) & 1.1(3) & A & 6.7(8) & 0.94(14) & A & 6.4(5) & 1.00(7) & ? \\
S60 & 8.8(5) & 0.8(2) & ? & 6.8(12) & 1.1(4) & A & 8.0(10) & 0.9(2) & ? & 5.8(3) & 0.84(16) & A\\
S63$^{f}$ & $\dots$ & $\dots$ & $\dots$  & 3.1(6) & 3.3(8) & A & $\dots$ & $\dots$ & $\dots$ & 6.4(5) & 2.84(16) & A\\
\noalign{\smallskip}
\hline
\noalign{\smallskip}
\end{tabular}
\\$^{a}$ Previously classified as Type A by \citet{bok03}, using different good times, frequency ranges and PDS classifications.
\\
$^{b}$Located within a globular cluster\\
$^{c}$Chandra resolved S47 into two sources; the information presented here is based on combined data from Chandra and XMM-Newton observations \citep{bko04}\\
$^{d}$Identified as a possible black hole by \citet{will04}\\
$^{e}$The high spectral indices presented in this table do not indicate the best fit models.\\
$^{f}$S63 was not in the EPIC field of view for Observations 1 and 3.  
\end{table*}

 Van der Meer et al. (2005) measured the  mass of the primary in SMC X-1 to be 1.05$\pm$0.09 M$_{\odot}$, using high resolution radial velocity measurements, and assuming that the secondary filled its Roche lobe. 
\citet*{vnq05} also measured the mass of the neutron star in SMC X-1. They obtained a maximum neutron star mass by assuming a Roche lobe filling secondary, and also obtained a minimum mass by assuming an inclination of 90$\degr$. In each case they corrected for X-ray heating of the secondary, unlike  \citet{vdm05}. \citet{vnq05} obtained upper and lower mass limits for the neutron star of 1.01$\pm$0.10 and 0.73$\pm$0.08 M$_{\odot}$ respectively before correction; this upper limit agrees well with the mass found by  \citet{vdm05}. After correction, maximum and minimum masses of 1.21$\pm$0.10 and 0.91$\pm$0.08 M$_{\odot}$ were obtained.
 Hence, the low luminosity Type B and flat PDS observed in S4 may be due to a low mass primary.

Alternatively,  it is possible that S4 is hysteretic, with the transition from Type B to Type A occurring at a lower luminosity than the transition from Type A to Type B, like the Galactic neutron star LMXB Aql X-1 \citep[see e.g.][]{mac03}. 
}

\subsection{X-ray sources associated with supernova remnants}
\label{snr}
Two sources, S38 and S58, are associated with SNRs, but also  exhibit Type A PDS.  In fact, the SNR of S58 has been resolved by Chandra \citep{kong02}. 
Only a handful of accreting compact objects have been associated with SNRs:  SS\thinspace 433 \citep[e.g.][]{dub98}, MF\thinspace 16 in the spiral galaxy NGC 6946 \citep[e.g.][]{rc03}, XTE\thinspace J0111.2$-$7317 in the SMC\citep{chr00} and RX\thinspace J050736$-$6847.8 in the LMC \citep{chu00}. None of these are thought to be LMXBs.

 Monte-Carlo simulations of LMXB formation  have shown that neutron star LMXBs have an average systemic velocity of 180$\pm$80 km s$^{-1}$ \citep[see e.g.][]{bp95}. The X-ray remnant of  S58 is ~40 pc in diameter, and \citet{kong02} estimate the age of the remnant to be $\sim$3,000-20,000 yr. Hence, the neutron star is unlikely to have  travelled further than $\sim$10 pc. The extraction region for S58 has an angular size of 40$^{''}$, corresponding to $\sim$150 pc; hence any LMXB formed in the supernova would be within our extraction region. We  further discuss S58 in a separate paper \citep{will05}.

\subsection{The  black hole candidates}
\label{bhc}

{ We found 12 sources that exhibited Type A variability at 0.3--10 keV luminosities $>$4$\times$10$^{37}$ erg s$^{-1}$ in at least one observation. In addition, the PDS of S14 were previously identified as Type A in Observations 1, 3 and 4 by \citet{bok03}, using different good time intervals and frequency ranges; also S51 was identified as a black hole candidate by \citet{will04} from its transient behaviour.  All these sources  are seen as a single point source by Chandra.

}
 The { 14 systems} are listed in Table~\ref{lspec}; the luminosity, spectral shape and PDS type are given for each of the four XMM-Newton observations  of each system. The spectral shape is illustrated by the photon index of the best fit power law model. 

 We have already published results for two of our black hole candidates.
S14 is the prototype for our new method for identifying black hole candidates, and is discussed in \citet{bok03}. S47 is a point source in the XMM-Newton observations, but was resolved by Chandra into two sources; we identified the northern source as a black hole candidate after combining data from these XMM-Newton observations with 35 Chandra observations \citep{bko04}.

\subsubsection{Could the  high luminosity Type A { variability} be due to beaming?}
\label{beaming}
{ 
Even if we assume that $l_{\rm c}$ $\sim$0.1,  systems that exhibit Type A variability at high luminosities could contain a neutron star, or even a white dwarf, if the emission were highly beamed or anisotropic. Such beaming could arise from the relativistic boosting of jet emission,  or from reprocessing of X-rays by a warped inner disc \citep[c.f.][]{pring96,fhm06}. Alternatively, the disc could simply have a lower optical depth to scattering in some directions(e.g. along the rotational axis) than in others \citep[see e.g.][]{king01}.

\citet{nm05} studied 10 Galactic black hole candidates to determine whether their quiescent luminosities were dependent on their inclinations; i.e., whether the quiescent emission was significantly beamed. However, they found no significant trend to suggest that relativistic beaming was significant. They compared the observed luminosities and inclination angles with beaming models for different Lorentz factors, $\gamma$, and found that $\gamma$ $<$1.24 at the 90\% confidence limit, if the X-ray emission originates from a jet. 
Similarly, \citet{m03} found that $\gamma$ $\la$1.4 in the X-ray emitting gas in GRO\thinspace J1655$-$40 during the low/hard state. { With such low Lorentz factors observed, we therefore conclude that the high luminosity Type A variability is unlikely to be due to relativistic Doppler boosting.}

{ We cannot rule out the possibility that any of our sources are emitting anisotropically. Assuming $l_{\rm c}$ $\sim$0.1, and a 1.4 M$_{\odot}$ primary, we define a hypothetical beaming factor for a source, $b_{\rm h}$, as the highest luminosity for Type A variability in the source, divided by 2$\times$10$^{37}$ erg s$^{-1}$. If the anisotropic emission were associated with the disc, then it would be feasible for the beam to precess. Hence, a source that varies greatly between observations may be emitting in precessing, anisotropic beam. S56 is one such source, and is further discussed in Sect.~\ref{sss}.

}

\subsubsection{X-ray properties of the black hole candidates}

{ 
Thirteen out of the fourteen sources { listed in Table~\ref{lspec}} have SEDs consistent with Galactic black hole LMXBs. However, S56 is a supersoft source, with a temperature of $\sim$20--30 eV, and cannot be considered as a candidate black hole LMXB in the low/hard state. We discuss the possible nature of S56 in Sect.~\ref{sss}. We  now briefly describe the properties of each black hole candidate  in turn.

{$\bullet$ S7~~~}We observed Type A variability from S7 at a 0.3--10 keV luminosity of 8$\pm$2$\times 10^{37}$ erg s$^{-1}$ in Observation 4. However, S7 exhibited Type B variability at 3.7$\pm$0.3 and 5.3$\pm$0.3  $\times 10^{37}$ erg s$^{-1}$ in observations 1 and 3 respectively. { The power law component of the SED of S7 in Observation 4 has a spectral index of 2.47$\pm$0.12; hence, S7 appears to be in the steep power law state, as defined by \citet{mr03}. }

{ Assuming $l_{\rm c}$ $\sim$0.1, then the low luminosity Type B variability observed in Observation 1 indicates that the primary is relatively low mass. However}, since S7 was not seen in either the Einstein or ROSAT surveys, it could be a recurring transient, and in decay from outburst during Observations 1 and 3. In this case, the transition from { Type B to Type A} during the decay of the outburst would occur at a significantly lower luminosity than $L_{\rm c}$.  This scenario would, however, require at least two outbursts { in the 18 months} between Observation 1 and Observation 4.

If the emission from S7 were beamed, then the beaming factor $b_{\rm h}$  would be $\ga$3 if the primary is 1.4 M$_{\odot}$ neutron star. The brightest observation of S7 is only $\sim$50\% brighter than the faintest, so we don't consider S7 to be a strong candidate for beaming.

{$\bullet$ S12~~~}Observation 4 revealed Type A variability in S12 at a luminosity of 6.0$\pm$0.3 $\times 10^{37}$ erg s$^{-1}$. The other observations of S12 had luminosities 4.6--5.0$\times 10^{37}$ erg s$^{-1}$, but no PDS were classified.  However, all the SEDs from S12 were consistent with { low/hard state} SEDs (i.e., consistent with $\Gamma$ $\sim$1.5--2.1). 

{$\bullet$ S14~~~}
There were three differences in the analysis of S14 { between the current paper and  \citet{bok03}}. Firstly, we were able to use the whole of Observation 1, { rather than a shorter good time interval},  as the background flaring did not significantly contribute to the 0.3--10 keV flux. Secondly, our PDS were constructed from 512 bins of 2.6 s. Finally, and most importantly, we allowed $\alpha$, the spectral index of the PDS below the break frequency, to vary. In observations 1, 3 and 4, $\alpha$ was found to be 0.11$\pm$0.01, 0.24$\pm$0.03 and 0.27$\pm$0.04 respectively; hence the PDS  were not well fitted by models where $\alpha$ = 0. In addition, the breaks occurred at high frequencies, so that the PDS were difficult to classify using frequencies $<$0.1 Hz.

 We showed in \citet{bok03} that a two component model was required to fit the Observation 4 SED of S14, comprising a 1.2--2.0 keV blackbody/disk blackbody component and a power law component with $\Gamma$ = 2.4--3.4. We argued incorrectly that the SEDs of Observations 1 and 3 must also have two components; while the SEDs from Observations 3 and 4 are very similar in shape and flux, the SED of Observation 1 is well described by a single  power law with $\Gamma$ $\sim$1.7. We now conclude that S14 was in the low/hard state in Observation 1 and the very high state in Observations 3 and 4.

{$\bullet$ S19~~~}Type A variability was observed at 9.7$\pm$0.5$\times 10^{37}$ erg s$^{-1}$ in Observation { 4} , while Type B variability was observed at 3.3$\pm$0.3 and 5.1$\pm$0.5 $\times 10^{37}$ erg s$^{-1}$ in Observations 1 and 3 respectively.  The SED for Observation 4 is fitted by $\Gamma$ = 2.47$\pm$0.09, which is consistent with the very high (steep power law) state. However, the low Type B luminosity in Observation 1 indicates a { low primary mass, perhaps  $\sim$3 M$_{\odot}$} if $l_{\rm c}$ $\sim$0.1. 

{$\bullet$ S37~~~}We see Type A variability from S37 in Observations 3 and 4,  at luminosities of 8.6$\pm$0.3 and 8.7$\pm$0.3 $\times 10^{37}$ erg s$^{-1}$ respectively. The SEDs of these observations fit with $\Gamma$ $\sim$1.8--1.9, consistent with the low/hard state. A flat PDS is observed in Observation 2, at a luminosity of 8.9$\pm$0.7$\times 10^{37}$ erg s$^{-1}$, while the PDS from Observation 1 is unclassified.  S37 is therefore consistent with being a canonical black hole.

{$\bullet$ S40~~~}Type A variability is seen in Observation 3 with a SED fitted by  $\Gamma$ = 1.88$\pm$0.09, at a luminosity of 4.9$\pm$0.3$\times 10^{37}$ erg s$^{-1}$. Additionally, a flat PDS was observed in Observation 4 at 4.3$\pm$0.2$\times 10^{37}$ erg s$^{-1}$. S40 is therefore consistent with having a black hole primary, with a relatively low mass. 

{$\bullet$  S43~~~}Observations 2 and 3 revealed Type A variability with SEDs consistent with the low/hard state at luminosities of 5.1$\pm$0.6 and 5.5$\pm$0.3 $\times 10^{37}$ erg s$^{-1}$ respectively.  In addition, S43 had a flat PDS in Observation 1, at a luminosity of 4.8$\pm$0.4$\times 10^{37}$ erg s$^{-1}$. However, S43 is associated with a globular cluster (see Sect.~\ref{globclust}). The maximum variation in luminosity $\sim$20\%, while $b_{\rm h}$ $\ga$3; it may be beamed, but we have no evidence to show it.

{$\bullet$ S47~~~}Chandra resolves S47 into two sources. The northern source is transient, and contains the black hole candidate, while the southern source is associated with a globular cluster. \citep[see][]{bko04}.

 {$\bullet$ S51~~~}\citet{will04} identified S51 as a black hole candidate, from its transient behaviours observed in Chandra and HST observations. Our XMM-Newton observations  reveal little extra information about the system. S51 exhibits Type A variability in Observation 3, hence it is probably disc accreting. However, the 0.3--10 keV luminosity derived from our best fit to the SED is only 4$\pm$4$\times$10$^{36}$ erg s$^{-1}$, { so we cannot use high luminosity Type A variability to classify it is a black hole candidate. }

{$\bullet$ S53~~~}In Observations 1--3, S53 is barely detected, and the PDS are unclassified. In Observation 4, a Type A PDS is observed at 4.2$\pm$0.2$\times 10^{37}$ erg s$^{-1}$, while the SED is fitted with $\Gamma$ = 1.83$\pm$0.08, consistent with the low hard state. We therefore consider S53 to be a M31 counterpart to  the Galactic black hole transients. 

{$\bullet$ S59~~~}Type A variability was seen in Observations 2 and 3, at luminosities of 9.1$\pm$1.5 and 6.8$\pm$0.7 $\times$10$^{37}$ erg s$^{-1}$ respectively. A flat PDS was observed at 7.8$\pm$1.0$\times$10$^{37}$ erg s$^{-1}$ in Observation 1, consistent with our black hole model. However, the SEDs are fitted with $\Gamma$ $\sim$0.9--1.1; i.e. the SEDs are too hard for the low/hard state in LMXBs. Instead, S59 could be a black hole HMXB. 

{$\bullet$ S60~~~} The SEDs of S60 are also too hard for LMXBs in the low/hard state: $\Gamma$ $\sim$0.8--1.1. Type A variability was observed at luminosities of 6.8$\pm$1.2 and 5.8$\pm$0.3 $\times$10$^{37}$ erg s$^{-1}$; no other PDS were classified. S60 may also be a HMXB.

{$\bullet$ S63~~~} The source was only in the field of view in Observations 2 and 4. Type A variability was seen in both, at luminosities of 3.1$\pm$0.6 and 6.4$\pm$0.5 $\times$10$^{37}$ erg s$^{-1}$ respectively.  However, the SEDs of S63 in Observations 2 and 4 are too soft for low/hard state spectra, suggesting that we are observing S63 in the steep power law state. The steep power law state is sometimes observed at the transition from Type A to Type B; hence, one of these luminosities may correspond to $L_{\rm c}$ in the 0.3--10 keV band. If S63 is hysteretic, like the Galactic black hole LMXBs, then $L_{\rm c}$ could be 6.4$\pm$0.5$\times$10$^{37}$ erg s$^{-1}$. Otherwise, $L_{\rm c}$ could be as low as 3.3$\pm$0.8$\times$10$^{37}$ erg s$^{-1}$; this lower $L_{\rm c}$ would indicate a primary mass of $\sim$3 M$_{\odot}$.
}

}

\subsubsection{A new population of persistently bright black hole LMXBs?}
Perhaps the most surprising result of this survey is that only { 3 out  of the 13} black hole candidates are transient, and  8 of the remaining 10 are  most likely LMXBs. By contrast, all of the confirmed Galactic black hole LMXBs are transient.
{ 
{ Indeed  S14, S19, S40, S59 and S63} are present in the 1979--1980 Einstein surveys \citep{tf91}, the 1991 and 1992 ROSAT surveys (S97, S01), as well as all pointed Chandra ACIS (W04) and XMM-Newton observations.  The SED of S59 suggests it may be a HMXB, but the other four sources appear to be black hole candidates in persistently bright LMXBs. }

{ Alternatively, these} sources could instead be exhibiting long duration, $\sim$20 year, outbursts, similar to Galactic black hole LMXBs GX 339$-$4 \citep[e.g.][]{zdz04} or GRS 1915+105 \citep[e.g.][]{bel00}. However, the first extra-solar X-ray source was  discovered  only $\sim$40 years ago \citep{gia62}; hence one might well ask if any LMXB is truly persistent. 

 The existence of persistently bright LMXBs is in fact predicted by theory. {  Evolutionary theory predicts a substantial population of short period black hole LMXBs with unevolved companions that would exceed the critical mass transfer rate for stable disc accretion \citep[see ][ and references within]{kks97}}. Hence large numbers of persistently bright black hole LMXBs would be  expected.

Disc accretion at low mass transfer rates is unstable; { however, in LMXBs, disc irradiation plays a role} \citep{vp96,kkb96}. The critical accretion rate that stabilises the disc, $\dot{M}_{\rm crit}^{\rm irr}$, is proportional to $R_{\rm D}^{2}$, where $R_{\rm D}$ is the disc radius; hence, $\dot{M}_{crit}^{irr}$ $\propto$ $P^{4/3}$ by Kepler's laws, where $P$ is the orbital period.  The fact that Galactic black hole LMXBs are transient where neutron star LMXBs with a similar period are persistent { suggests} that irradiation is weaker in black hole LMXBs \citep{kks97}.

Using the assumptions given in \citet{kks97}, 
\begin{equation}
\dot{M}_{\rm crit}^{\rm irr} \simeq 2.86\times10^{-11} m_1^{5/6} m_2^{-1/6} P_{\rm h}^{4/3} {\rm M_{\odot} yr^{-1}}
\end{equation}
where $m_1$ is the primary mass  and $m_2$ is the secondary mass in solar units, and $P_{\rm h}$ is the orbital period in hours { \citep[see also][]{dub99}}. \citet{kks97} calculated $\dot{M}$ over $P_{\rm h}$ $\sim$1--8 hr for black holes with masses 2, 5 and 10 M$_{\odot}$, for comparison with $\dot{M}_{\rm crit}^{\rm irr}$. Both the 2 and 5 M$_{\odot}$ black holes were sufficiently irradiated for a stable disc in the range $\sim$3--8 hr. For the 10 M$_{\odot}$ black hole, $\dot{M}$ $<$ $\dot{M}_{\rm crit}^{\rm irr}$ for all periods. Hence, LMXBs with lower mass primaries are more likely to be persistent. However, due to uncertainties in their assumptions, \citet{kks97} acknowledge that even 10 M$_{\odot}$ black holes could be persistent in this range of periods, { while long-period LMXBs are always expected to be transient \citep{king00}.}

\begin{figure}[!b]
\resizebox{\hsize}{!}{\includegraphics[angle=270]{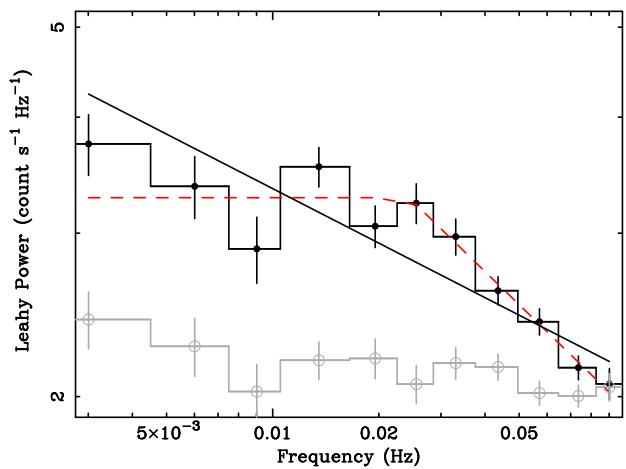}}
\caption{  PDS of S53 from Observation 4, using 192 intervals of 64 bins. We show the best fits to the PDS using power law (solid) and broken power law (dashed ) models. For the best fit power law and broken power law models,  $\chi^2$/dof = 35/9 and 11/8 respectively; the broken power law is formally acceptable, while the power law is not. We also show the PDS of the background, in grey with open circles. It is clear that the source PDS is not dominated by background variability.
}\label{bp}
\end{figure}

\subsubsection{Multi-wavelength observations of the black hole candidates}

{ Of the thirteen black hole candidates, only S43 and S53 have  been observed at other wavelengths. S43 is associated with the globular cluster Bo 144 \citep{bat87}, and has been observed in UV the optical and infrared \citep{gal04}; however, the counterpart to S43 will be lost in the cluster, so we will discuss S43 no further.}
\citet{will04} report results from Chandra/HST observations of S53, using the designation ``r3-16''. 
They  find a { roughly} persistent U band counterpart to r3-16 that resembles an unresolved binary; the U magnitude was measured to be 20.82$\pm$0.06 in 2001, August and 2001,  December, and 21.11$\pm$0.02 in 2002, January. Measurements from the WFPC2 instrument on board HST suggest a separation { between the two components} of 0.82$\pm$0.08 pc, if located in M31. However, ground based  measurements using the  B and V band data from the Local Group Survey of  \citet{mas01} yield a separation of 3.9$\pm$0.6 pc. 
 \citet{will04} suggest that S53 could be either a blend of foreground stars, a  cataclysmic variable (CV) at a distance  $\la$1 kpc, or a background AGN. They note that the X-ray SED is harder than usual for a CV, and that the SED and line-of-sight absorption are more in keeping with a background AGN. 
However, the Type A PDS exhibited by S53 in Observation 4 means that it cannot be a background AGN,  and is unlikely to be a blend of foreground stars. 

Furthermore, the 37W radio survey of M31 \citep{walt85} revealed a 1.1 mJy radio source (37W153) just { 0.6}$\arcsec$ from the Chandra position of S53; the survey was conducted at a frequency of 1412 MHz and  had a 23$\arcsec$ $\times$ 35$\arcsec$ resolution. Radio emission is common in black hole LMXBs in the low state \citep[e.g.][]{fen04}, but is rare in CVs. Claims have been made for detecting radio emission from several magnetics CVs \citep[e.g.][]{chan87,pav94}, as well as in  novae, dwarf novae, and recurrent novae \citep{chan87}. However, the only confirmed radio detection is from the peculiar magnetic CV AE\thinspace Aquarii \citep[e.g.][]{mei03}.   

{ Cygnus X-3 is one of the brightest X-ray binary radio sources; it exhibits flares up to $\sim$20 Jy \citep[see e.g.][]{walt91}. \citet{pbp00} estimate the distance of Cygnus X-3 to be $\sim$9 kpc, using Chandra observations of its scattering halo. If we place Cyg X-3 in M31, the radio flares can reach $\sim$2.5 mJ. Hence, it is feasible for S53 to be a M31 analogue to Cyg X-3.}

 S53 is unlikely to be a magnetic CV since its Type A variability (Fig.~\ref{bp}) indicates disc accretion. 
 \citet{will04} found no clear correlation between the X-ray flux of S53 and the U band flux of its counterpart; hence, the association between the UV and X-ray sources may be coincidental.

{
\subsection{ The supersoft sources}
\label{sss}

There are four sources in the survey that have been classed as supersoft: S35, S56, S57 and S58. Many supersoft sources are thought to be accreting white dwarf binaries, with sufficiently high mass transfer rates for stable hydrogen burning \citep{vdh92}. However, S58 is a supernova remnant, and discussed in Sect.~\ref{snr}.

Blackbody radiation models assume that the radiation source is optically thick at all energies. However, the opacities of model white dwarf atmospheres are dominated by only a few ions in the temperature range typical of supersoft sources; as a result, the  X-ray spectra deviate significantly from blackbody emission \citep{htk94}. 
A blackbody fit to a white dwarf atmosphere  will significantly over-estimate low-energy flux and radius \citep{htk94}.
Indeed, blackbody fits to the spectra of S35 and S56 could be formally rejected in Observations 3 and 4, with best fit $\chi^2$/dof $>$2.

Detailed modelling of the supersoft sources with white dwarf atmosphere models is beyond the scope of this paper. However, neutron star atmosphere (NSA) models give qualitatively similar SEDs, so we applied NSA models to the supersoft sources as a first approximation.

S35, S56 and S57 have broadly similar characteristics; all are well described by a NSA model, with temperatures of 20--30 eV, and apparent 0.3--10 keV luminosities $>$10$^{38}$ erg s$^{-1}$ in at least one observation. However, S57 is a transient that exhibited $\sim$850 s pulsations in Observation 1 \citep{osb01}, while the 0.3--10 keV luminosity of S56 apparently varied  over 1.5--40 $\times$10$^{37}$ erg s$^{-1}$, exhibiting the highest luminosity Type A variability in the survey.  

The pulsations exhibited by S57 shows that the emission is anisotropic. Furthermore, S56 exhibited Type A variability at $\sim$4$\times$10$^{38}$ erg s$^{-1}$, suggesting $b_{\rm h}$ $\ga$20; S56 also varied by a factor of $\sim$20 between observations, { indicating that the emission is likely to be anisotropic}. Hence we could expect the emission from  S35 to be  anisotropic also.
Indeed, collimated jets have been observed from the Galactic supersoft sources RX\thinspace J0925.7$-$4578 \citep{mot98}, RX\thinspace J0513.9$-$6951 \citep{cram96,south96} and RX\thinspace J0019.8+2156 \citep{tom98,bec98}. { It is therefore quite possible that  S35, S56 and S57 are} all disc accreting white dwarf binaries that exhibit some sort of jet.

\section{Summary}
\label{summary}
{  We have conducted a survey of PDS and SEDs obtained from 4 XMM-Newton  observations of the central region of M31. The sample consisted of the 63 X-ray sources that exceeded a 0.3--10 keV EPIC-pn intensity of 0.02 count s$^{-1}$ in at least one observation.} 
 Four XMM-Newton sources were identified as multiple sources by Chandra: S22, S31, S33 and S47; in \citet{bko04}, we show that the northern source in S47 is a possible black hole.  { A further four sources are supersoft sources. }The known foreground star, S6, showed neither Type A nor Type B PDS, as expected. The remaining { 54} sources are consistent with being LMXBs; {  we identify 13 as black hole candidates.}

Variability over a 100 s time scale is observed in 55 { out of the 63}  sources. Of these, { 40} sources show Type A variability, meaning that they cannot be background AGN and are most likely to be LMXBs; a further { 6} show Type B but not Type A PDS, making them likely accreting binaries. Hence, we identify { a total of 46} likely  X-ray binaries out of 63 X-ray sources, from their PDS and SEDs.  We find that the PDS and SEDs correspond very well to the spectral and timing states of Galactic X-ray binaries; 2 sources appear to be HMXBs while the other 44 behave like LMXBs.

   \citet{pie05} conducted a survey of the X-ray population in all archival XMM-Newton observations of M31. They detected 856 point sources, including 7 X-ray binaries and 9  candidate X-ray binaries. They also reported 27 X-ray source associations with globular clusters, with 10 further possible associations; these are also likely LMXBs. Of these 53 sources, 21 are in our sample and 10 of those exhibited Type A PDS. Hence, 36 of our X-ray binaries are newly identified in this work.

 In addition, luminosity variations greater than 5$\sigma$ between observations were exhibited by 57 of the sources. Of these, 7 are transient, in that they vary in intensity by a factor of $\ga$10, and are absent in at least one observation. The fraction of variable X-ray sources is 90\% in our survey, higher than the 50\% quoted by K02 for the field population;   this is simply because the XMM-Newton observations were deeper than the ones performed by Chandra.
 \citet{tru04} analysed XMM-Newton observations of M31 X-ray sources  in globular clusters, and found 80\% to be variable. They suggested that the variability for LMXBs in globular clusters was higher than those in the field,  comparing  XMM-Newton observations of globular cluster X-ray sources with Chandra observations of the general population. Our XMM-Newton observations suggest that the globular cluster X-ray sources are no more variable than the general population.

 Among our  $\sim$50 likely LMXBs, we have observed 1 Z-source, 2 dippers, 7 transients and 13 candidate black holes. Out of the $\sim$150 Galactic LMXBs, 6 or 7 are Z-sources, $\sim$10 are dippers, $\sim$70  are transients and 37 are black hole candidates \citep[e.g.][]{lvv01}. Hence, our sample contains $\sim$half the number of Z-sources and dippers, and $\sim$ one third of the transients expected from the Galactic population. However, we found a comparable number of black hole candidates.

\section*{Acknowledgments}
This work was funded by PPARC, and used publicly available data from the Chandra Archive mirror at LEDAS (at http://ledas-www.star.le.ac.uk/chandra).  We thank W. Pietsch for providing the list of sources classified as X-ray binaries or globular clusters in the \cite{pie05} survey.  We also thank Fill Humphrey for the use of his {\sc Filltools} software suite. R.B. would like to thank W. Clarkson, R. Cornelisse, T. Maccarone, J. Wilms and D. Gelino for useful discussions.   We would also like to thank the anonymous referee for their thorough and constructive comments that considerably improved the paper.

\appendix

\section{Survey results in detail}
\label{appendix}

{ \begin{table}[!b]
\centering
\caption{ Limits in $\chi^2$ corresponding to probabilities of 0.001 and 0.05, for a  given number of degrees of freedom}\label{chidat}
\begin{tabular}{lllll}
\noalign{\smallskip}
\hline
\noalign{\smallskip}

D.O.F & $\chi^2$ ($H_0$ = 0.001) & $\chi^2$ ($H_0$ = 0.05)\\
\noalign{\smallskip}
\hline
\noalign{\smallskip}
8 & 26 & 16 \\
9 & 28 & 17\\
11 & 31 & 20\\
12 & 33 & 21\\
\noalign{\smallskip}
\hline
\noalign{\smallskip}

\end{tabular}
\end{table}

For each observation of every source, we recorded the fractional r.m.s variability, $F_{\rm var}$ \citep{vgn03}, PDS characteristics and SED characteristics. 

We obtained best fit parameters to the 64- and 128- bin PDS using power law and broken power law models, and used F-testing to ascertain the probability that the broken power law is the better fit.  The null hypothesis probability, $H_0$, is the probability that the deviation of observed data from the model is solely due to random statistical fluctuations.  We rejected any fits where $H_0$ $<$0.001, and considered any model with $H_0$ $>$0.05 an acceptable fit. 
Power law fits to our 64 and 128 bin PDS have 9 and 12 degrees of freedom, respectively, while broken power law fits have 8 and 11 degrees of freedom for our 64 and 128 bin PDS. We present the $\chi^2$ corresponding to $H_0$ = 0.001 and 0.05 for 8, 9, 11 and 12 degrees of freedom in Table~\ref{chidat}. For example, we reject fits with $\chi^2$/dof $>$26/8 and accept fits where $\chi^2/$dof $<$16/8.

  We fitted each SED with the following simple { spectral} models: blackbody (BB), bremsstrahlung (BR) and power law (PO), all suffering line-of-sight absorption. In some cases, a second component was required; in this case, we used a blackbody + power law (2C) model. In addition, the supersoft sources were fitted with neutron star atmosphere (NSA) models; this was to represent a thermal model with a sharper drop off at low energies than a blackbody.

We present the properties of each source in Table~\ref{allprops}; we present the results of each observation on a different line. Numbers in parentheses indicate uncertainties in the last digit.  Column 1 is the source name. Column 2 is the  fractional  r.m.s variability of the background subtracted lightcurve. Columns 3--7 describe the fits to either the 64- or 128 bin PDS; we present results for the PDS with the lowest F-test probability, to give the most conservative detections of Type A variability.  Columns 3 and 4 are the spectral index, $\gamma$ and $\chi^2$/dof for the power law model. Similarly, columns 5, 6 and 7 are the break frequency, $\nu_{\rm c}$, spectral index for $\nu$ $>$  $\nu_{\rm c}$, $\beta$, and $\chi^2$/dof for the broken power law model. Uncertainties in columns 3--7 are at the 1 $\sigma$ level. The best fit model is indicated in boldface.   Column 8 gives the F-test function, $F$, which is the probability that the improvement in fitting a broken power law to the PDS over a power law is due to chance. Column 9 is the PDS type. Columns 10--14 describe the SED. Column 10 gives the best fit spectral model. Column 11 is the equivalent line-of-sight absorption, $N_{\rm H}$, in units of 10$^{22}$ H atom cm$^{-2}$.  Column 12 describes the shape of the SED, in terms of the temperature, k$T$, in keV units for thermal models (BB, BR or NSA), or the photon index, $\Gamma$, for the PO or 2C models. Column 13 gives the $\chi^2$/dof for the best fit to the SED, while S14 gives the 0.3--10 keV luminosity, in units of 10$^{37}$ erg s$^{-1}$. Uncertainties in columns 10--12 and 14 are given at the 90\% confidence level, and frozen parameters are indicated by ``f''. 

The PDS are classified as follows. If $F$ $<$0.001, the PDS is of Type A. If $F$ $>$ 0.001, but $H_0$ $<$0.001 for the best  power law fit and $H_0$ $>$0.05 for the broken power law fit, then we classify the PDS as Type A$^{\ast}$, and treat it as Type A. If, however, the PDS is of Type A but dominated by variability in the background, then we denote it as Type A$^{\dagger}$ in Table~\ref{allprops}, and treat it as unclassified. If a power law gives $\chi^2$/dof $\la$1.0, and $\gamma$ is $>$3$\sigma$ above zero, then the PDS is of Type B. If the PDS does not rise significantly above a power of 2, then we classify it as a flat PDS, indicated by $-$. Any PDS not classified as above is unclassified, indicated by a question mark.
}

 \begin{table*}[!t]   
\renewcommand{\baselinestretch}{.5}
\renewcommand{\tabcolsep}{1mm}
\caption{PDS and SED properties for each observation of every source. Observations 1--4 are described on separate lines. Numbers in parentheses indicate uncertainties in the last digit. Columns 1 and 2 give the source name and fractional r.m.s. variability. Columns 3--9 concern fits to the PDS; 3 and 4 give the spectral index and $\chi^2$/dof for the best fit power law model to the PDS, while 5, 6 and 7 show the break frequency in mHz, spectral index for $\nu$ $>$ $\nu_{\rm c}$ and $\chi^2$/dof for the best broken power law fit to the PDS; column  8  is  F-test probability that the improvement in fitting the broken power law  over the power law  is due to chance, and column nine gives the PDS Type. Uncertainties in fits to the PDS are given at the 1$\sigma$ level. Columns 10--14 describe fits to the SED; column 10 gives the best fit model (see text for abbreviations); 11 gives the equivalent Hydrogen absorption in the line of sight; 12 gives either the photon index, or temperature, appropriate to the model; 13 gives the $\chi^2$/dof for the best fit model and 14 gives the 0.3--10 keV luminosity. Uncertainties in fitting the SED are at the 90\% confidence level. ``$\dots$'' indicates that no good SED was obtained; either the source was not observed with the PN, or the SED had negative counts.}\label{allprops}
\begin{tabular}{llllllllllllllll}
\noalign{\smallskip}
\hline
\noalign{\smallskip}
& & \multicolumn{2}{c}{Power Law Fit} & \multicolumn{3}{c}{Broken Power Law Fit}& & &\multicolumn{5}{c}{SED FIT} \\
S & { $F_{\rm var}$} & $\gamma$ & $\chi^2_{\nu}$& $\nu_{\rm c}$ & $\beta$ & $\chi^2_\nu$& $F$  & Type & Spec& $N_{\rm H}$$^a$ & $\Gamma$/kT$^b$ & $\chi^2_\nu$ &   L$^c$\\
\noalign{\smallskip\hrule\smallskip}
1  &2(3) & 0.08(2) & 15/9 &  {\bf 53(8)} & {\bf 0.40(14)} & {\bf 8/8} & 0.03 & ?& PO & 0.1 f& 3.8(14)& 12/12 &  0.3(2)\\
& 0.2(3) &{\bf  0.10(3)} &{\bf 14/9} &44(1) & 0.33(13) & 15/8 & NA & ? & PO & 0.18(10) & 2.0(5) & 6/8 &  1.7(3)  \\
& 0.27(16) & {\bf 0.06(16)}  &  {\bf 7/9} & 7(7) & 0.06(2) & 7/8 & NA & $-$ & PO & 0.33(9) & 2.3(3)&  16/27 & 1.9(3) \\
\vspace{0.07in}
&2.3(5) & 0.21(2) & 109/9 & {\bf 27(3)} & {\bf 0.49(5)} & {\bf 18/8} & 2 E-3 & A & $\dots$  & $\dots$& $\dots$ & $\dots$& $\dots$\\

2  &0.12(17) & {\bf 0.04(2)} & {\bf 12/12} &  100(200) & 0.1(20) & 16/11&  NA &$-$ & PO & 0.26(4) &1.78(9)& 30/40 & 5.0(4)\\
 & 0.20(9) & 0.13(3) & 21/9 & {\bf 47(10)} & {\bf 0.7(4)} & {\bf 11/8} & 0.03 & ? &PO &  0.24(11) & 1.78(3) & 7/9 & 4.7(6)\\
& 0.05(30) & 0.06(2) &8/9 & {\bf 60(7)} & {\bf 0.06(2)} & {\bf 7/8} & 0.77 & $-$ & PO & 0.32(4) & 2.03(11) & 74/77 &  5.5(3)\\\vspace{0.07in}
& 0.13(6) & 0.06(2) & 8/9 & {\bf 31(9)} & {\bf 0.14(5)} & {\bf 4/8} & 0.02 & $-$& PO & 0.39(4) & 2.24(10) & 70/68 & 5.9(4)\\

3  & 0.44(14) & 0.04(2) & 19/12  & {\bf 47(2)} & {\bf 0.22(8)} & {\bf 15/11} & 0.13 & ? & PO & 0.7 f & 2.0(5)& 30/28 & 1.3(7) \\
& 0.50(10) & 0.22(3) & 14/9 & {\bf 37(7)} & {\bf 0.6(2) } & {\bf 6/8} & 0.01 & ? &PO&   0.9(5) & 2.1(6) & 10/12 &2.1(6) \\
& 0.69(10) & {\bf 0.02(2)} & {\bf 10/9 } & 57(2) & 0.69(12) & 11/8 & NA & $-$& PO & 0.7(3) & 1.7(6) & 91/80 & 1.5(4)  \\\vspace{0.07in}
&0.54(4) & 0.18(2) & 63/9 & {\bf 36(3)} & {\bf 0.54(7)} & {\bf 6/8} & 2 E$-$5 & A & PO & 0.6(2) & 2.0(3) & 33/34 & 2.0(2)  \\

4  & 0.3(22) & {\bf 0.06(2)} & {\bf 6/9}  & 30(20) & 0.10(7) & 8/8 & NA & $-$ & PO & 0.4(2) & 2.2(4) & 8/12 & 1.3(3)\\
& 0.4(2) & {\bf 0.12(3)} & {\bf 7/12}& 39(13) & 0.32(17) & 11/11 & NA & B & PO & 0.2(2) & 1.7(5) & 19/10 &  1.0(3)\\
& 0.6(2) & 0.17(2) & 47/9 & {\bf 28(3)} & {\bf 0.40(5)} & {\bf 12/8} & 1 E$-$3 & A$^{\dagger}$ & PO & 0.40(13) & 2.2(3) & 29/30 & 1.2(3) \\\vspace{0.07in}
& 0.40(10)& 0.19(2) & 86/9 & {\bf 36(3)} & {\bf 0.57(7)} & {\bf 9/8} & 4 E$-$5 & A$^{\dagger}$ & PO & 0.30(6) & 2.08(16) & 26/30 & 1.8(2)\\

5  & 0.7(3) & 0.13(2) & 26/12 & {\bf 25(6) } & {\bf 0.31(6) } & {\bf 10/11} & 2 E$-$3 & ? & PO & 0.5 f & 2.7(4) & 7/10 & 0.9(5)\\
&0.21(14) & 0.16(4) & 18/9 & {\bf 24(8)} & {\bf 0.35(10)} &{\bf  10/8} & 0.04 & ? &  PO & 0.5(2) & 2.0(4) & 13/17 &  3.3(5)\\
 & 0.38(15) & 0.16(2) & 46/9 & {\bf 44(5)} & {\bf 0.54(11)} & {\bf 39/11} & 0.27 & ? & PO & 0.3(2) & 2.0(4) & 34/31 & 1.2(2)\\\vspace{0.07in}
& 0.15(9) & 0.08(2) & 23/9 & {\bf 26(8)} & {\bf 0.16(5)} & {\bf 13/8} & 0.04 & ? & PO & 0.58(6) & 1.91(13) & 37/45 & 3.6(3)\\

6  & 0.02(7)& 0.06(2) & 27/9 & {\bf 57(7)} & {\bf 0.24(12)} & {\bf 23/8} & 0.28 & ? & BR & 0.49(3) &  1.89(17) &68/71& 10.0(10)$^{d}$ \\
& 0.03(20) & 0.05(4) & 16/12 & {\bf 70(60)} & {\bf 0.1(6)} & {\bf 15/11} & 0.80 & $-$ & BR & 0.55(8) & 1.5(3) & 12/20 & 8.9(9)$^{d}$ \\
& 0.06(7) & 0.04(2) & 13/12 & {\bf 11(14)} & {\bf 0.06(4)} & {\bf 12/11} & 0.44 & $-$ & BR & 0.49(3) & 1.91(12) & 167/137 & 10.5(5)$^{d}$  \\\vspace{0.07in}
& 0.11(3) & 0.14(2) & 30/9 & {\bf 33(5)} & {\bf 0.34(7)} & {\bf 3/8} & 5 E$-$5 & A$^\dagger$ & BR & 0.53(3) & 1.70(9) & 129/115 & 9.3(5)$^d$\\

7 & 0.15(9) & {\bf 0.07(2)} & {\bf 8/12}& 18(13) & 0.10(5) & 9/11 & NA & B & PO & 0.22(4) & 2.3(2) &  24/35 & 3.7(3)\\
& 0.06(13)& 0.17(3) & 15/9 & {\bf 36(9)} & {\bf 0.46(18)} & {\bf 13/8} & 0.33 & ? & PO & 0.16(7) & 2.0(3) & 9/13 & 4.9(5)\\
& 0.13(5) & {\bf 0.12(2)} & {\bf 7/9} & 20(8) & 0.18(6) & 6/8 & 0.29 & B & PO & 0.20(3) & 2.09(10) & 127/97 & 5.3(3) \\\vspace{0.07in}
&0.11(4) & 0.12(2) & 40/9 & {\bf 35(4) }& {\bf 0.38(6)} &{\bf 7/8} & 3 E$-$4 & A & PO & 0.28(8) & 2.47(12) & 142/105 & 8(2)\\

8 & 0.14(10) & 0.08(2) & 34/12 & {\bf 35(7)} & {\bf 0.32(8)} & {\bf 16/11} & 4 E$-$3 & A$^{\ast}$ &  PO & 0.40(8) & 2.4(2) & 28/27 & 3.3(3)\\
& 0.2(3) &{\bf  0.07(4)} & {\bf 14/12} & 16.5(8) & 0.10(5) & 14/11 & NA &? &  PO & 0.18(3) & 1.6(5) & 11/9 & 1.2(3)\\
& 0.38(16) & 0.12(2) & 19/9 & {\bf 33(7)} & {\bf 0.32(5)} & 8/8 & 0.01 & ? & PO  & 0.4(2) & 2.0(4) & 26/31 &1.73(17)  \\\vspace{0.07in}
& 0.36(11) & 0.12(2) & 36/9 & {\bf 36(5) } & {\bf 0.37(6)} & {\bf 12/8} & 4 E$-$3 & A$^{\ast}$ & PO & 0.3(2) & 1.7(2) & 15/22 & 1.3(2)   \\

9 & 0.10(11) &{\bf 0.14(2)} & {\bf 14/12} &  6(5) &  0.1(3) &  14/11 & 0.63 & $-^{2}$ & PO & 0.40(5) & 1/39(9) & 101/74 & 24.6(17)  \\
& $\dots$ &  $\dots$ &  $\dots$ &  $\dots$ &  $\dots$ &  $\dots$ &  $\dots$ &  $\dots$ &  $\dots$ &  $\dots$ &  $\dots$ &  $\dots$ &  $\dots$ & \\
&  0.07(14) &{\bf  0.13(2)} &{\bf  12/9 }& 60(60) &0.4(3) & 12/8 & NA & $-^2$ & PO & 0.49(5) & 1.48(9) & 158/149 & 23.5(14) \\\vspace{0.07in}
& $\dots$ &  $\dots$ &  $\dots$ &  $\dots$ &  $\dots$ &  $\dots$ &  $\dots$ &  $\dots$ &  $\dots$ &  $\dots$ &  $\dots$ &  $\dots$ &  $\dots$ & \\

10 & 0.13(5) & 0.12(2) & 18/12  & {\bf 54(6)} & {\bf 0.6(2)} & {\bf 12.11} & 0.04 & ? &  PO & 0.24(4) & 2.71(11) & 70/49 &  6.93(14)\\
& 0.1(2) & {\bf 0.08(3)} & {\bf 19/12} & 36.8(2) & 0.20(10)  & 20/11 & NA & ?&  PO  & 0.23(5) & 2.8(2) & 16/20 & 8.1(8)  \\
& 0.13(13) & 0.02(2) & 15/9 & {\bf 57(2) } & {\bf 0.27(11) } & {\bf 10/8} & 0.08 & $-$ & PO   & 0.26(5) & 3.1(3) & 45/65 & 4.4(4)\\\vspace{0.07in}
& 0.09(6) & 0.09(2) & 45/9 & {\bf 35(6)} & {\bf 0.28(6)} & {\bf 13/8} & 2 E$-$3 & A$^{\dagger}$ & PO & 0.29(2) & 3.14(11) & 124/89 & 7.8(5)\\

\noalign{\smallskip}
\hline
\noalign{\smallskip}
\end{tabular}
\\$^a$$N_{\rm H}$ is in units of 10$^{22}$ H atom cm$^{-3}$\\
$^b$Either spectral index or temperature, depending on the mode; kT is in units of keV\\
$^c$L is in units of 10$^{37}$ erg s$^{-1}$
\\$^{d}$These luminosities assume S6 is in M31, however it is a foreground star, hence its luminosity is reduced by $\sim$6 orders of magnitude.
\end{table*}

\setcounter{table}{1}
 \begin{table*}[!t]   

\renewcommand{\baselinestretch}{.5}
\renewcommand{\tabcolsep}{1mm}
\caption{  continued}
\begin{tabular}{llllllllllllllll}
\noalign{\smallskip}
\hline
\noalign{\smallskip}
& & \multicolumn{2}{c}{Power Law Fit} & \multicolumn{3}{c}{Broken Power Law Fit}& & &\multicolumn{5}{c}{SED FIT} \\
S & { $F_{\rm var}$} & $\gamma$ & $\chi^2_{\nu}$& $\nu_{\rm c}$ & $\beta$ & $\chi^2_\nu$& $F$  & Type & Spec& $N_{\rm H}$ & $\Gamma$/kT & $\chi^2_\nu$ &   L\\
\noalign{\smallskip\hrule\smallskip}
11 & 0.53(8) & 0.41(2) & 32/9  & {\bf 10(2)} & {\bf 0.52(4)} & {\bf 19/8} & 0.04 & ? & PO & 0.09(3) & 1.40(9) & 51/51 & 9.4(7)\\
&  0.03(24) & 0.10(3) & 20/9 & {\bf 21(12)} & {\bf 0.23(11)} & {\bf 6/8} & 0.07 & ? & PO & 0.10(5) & 1.33(13) & 28/21 &7.3(9)  \\
& 0.14(6) & 0.08(2) & 10/9 & {\bf 34(12)} & {\bf 0.21(10)} & {\bf 6/8} & 0.08 & $-$ & PO & 0.12(2) & 1.56(8) & 86/74 & 4.4(3)\\\vspace{0.07in}
& 0.10(3) & 0.13(2) & 27/12 & {\bf 22(4)} & {\bf0.22(4)} & {\bf 11/11} & 2 E$-$3 & ? & PO & 0.14(2) & 1.59(6) & 167/135 & 7.0(3) \\

12 & 0.04(9) & 0.10(2) & 22/12  & {\bf 42(7)} & {\bf 0.40(11)} & {\bf 10/11} & 3$\times$10$^{-3}$ & ? & PO & 0.38(6) & 2.21(15) &  45/43 & 4.8(5)\\
& 0.17(11) & 0.09(4) & 9/9 & {\bf 38(14)} & {\bf 0.31(16)} & {\bf 5/8} & 0.02 & ? &PO & 0.38(11) & 2.2(3) & 11/13 & 5.0(6)\\
&0.11(9) & 0.12(2) & 17/9 & {\bf 30(70) } & {\bf 0.22(5) }&{\bf  14/8} & 0.23 & ? & PO & 0.33(4) & 2.14(13) & 104/91 & 4.6(3)\\\vspace{0.07in}
& 0.15(4) & 0.17(2) & 32/9 & {\bf 29(3) } & {\bf 0.37(5)} & {\bf 5/8} & 2 E$-$4 & A & PO & 0.42(3) & 2.18(9) & 76/98 & 6.0(3)\\

13 & 0.18(12) &0.09(2) & 31/12 &  {\bf 40(6)} & {\bf 0.45(8)} & {\bf 18/11} & 0.02 & ? & PO & 0.13(6) & 1.8(2) & 25/24 & 1.7(2)\\
&0.28(18) & {\bf 0.11(4)} & {\bf 6/12} & 50(20) & 0.3(3) & 11/11 & NA & ? & PO &  0.19(15) & 2.6(8) &3/4 &1.6(2)  \\
&0.15(8) & 0.11(2) & 13/9 & {\bf 36(8)} & {\bf 0.34(10)} & {\bf 2/8} & 2 E$-$4 & A &PO &  0.13(3) & 1.85(14) & 62/71 & 2.6(2)\\\vspace{0.07in}
& 0.14(5) & 0.16(2) & 47/9 & {\bf 29(3) } & {\bf 0.37(5)} & {\bf 8/8} & 2 E$-$4 & A & PO & 0.16(3) & 1.99(9) & 95/75 & 3.3(4)\\

14& 0.06(7) & 0.05(2) & 8/9 & {\bf 30(20)}& {\bf 0.09(5)} & {\bf 7/8}& 0.40 & $-$ & PO & 0.19(2) & 1.73(8) & 85/81 & 7.7(4)\\
& 0.07(10) & 0.07(3) & 15/12 & {\bf 50(20)} & 0.3(3) & {\bf 13/11} & 0.21 & ? & PO & 0.13(7) & 1.7(2) & 9/11 & 5.7(7) \\
&0.11(2) & {\bf 0.13(2)} & {\bf 23/12} & 32(8) & 0.28(9) & 26/11 & NA & ? &2C & 0.22(10) &2.39(7) & 376/347 & 24.6(12)  \\\vspace{0.07in}
& 0.099(14) & 0.18(2) & 15/12 & 19(3) & 0.26(3) & 14/11 & 0.40 & ? & 2C & 0.23(2) & 2.33(6) & 373/375 & 23.8(5)\\

15 & 0.18(13) & {\bf 0.12(2)} & {\bf 9/9} & 29(8) & 0.21(7) & 10/8 & NA & B & PO & 0.32(9) & 2.0(3) & 28/20 & 1.9(3)\\
&0.1(4) & {\bf 0.11(4)} & {\bf 7/9} & 50(20) & 0.3(3) & 9/8 & NA & ? & PO & 0.17(11) & 2.4(5) & 4/8 &  1.2(3)\\
& 0.17(16)& 0.08(2) & 18/12 & {\bf 37(2)} & {\bf 0.26(6)} & {\bf 11/11} & 0.02 & ? & PO & 0.36(8) & 2.3(2) & 43/43 & 1.9(3)\\\vspace{0.07in}
& 0.33(9) & 0.17(2) & 22/9 & {\bf 29(4) } & {\bf 0.33(5)} & {\bf 9/8} & 7 E$-$3 & ? & PO & 0.39(8) & 2.3(2) & 17/25 & 1.4(2)\\

16 & 0.64(8) & 0.14(2) & 15/9  & {\bf 32(6)} & {\bf 0.32(7)}  & {\bf 8/8} & 0.02 & ? & PO & 0.2(2) & 2.1(7) & 38/33 &  1.5(4)\\
& 0.4(3) & {\bf 0.13(4)} & {\bf 3/9} & 28(14) & 0.23(12) & 5/8 & NA & B &PO & 0.2(2) & 1.8(7) & 4/8 & 1.4(4) \\
& 0.59(14) & 0.15(2) & 26/12 & {\bf 51(6)} & {\bf 0.63(17)} & {\bf 24/11} & 0.38 & ? & PO & 0.4(2) & 2.6(6) & 37/42 & 1.9(5) \\\vspace{0.07in}
& 0.30(11) & 0.15(2) & 33/9 & {\bf 25(4) } & {\bf 0.2(3)} & {\bf 5/8} & 1 E$-$4 & A & PO & 0.28(9) & 1.8(2) & 35/33 & 1.8(2)\\

17 & 0.14(9) &  0.08(2) & 14/9  & {\bf 25(8)} & {\bf 0.16(4)} & {\bf 11/8} & 0.23 & ? & PO & 0.10(4) & 1.53(14) & 35/31 & 3.4(3)\\
 & 0.02(76) & 0.03(3) & 13/12 & {\bf 50(30)} & {\bf 0.2(3)} & {\bf 13/11} & 0.78 & $-$ &PO &  0.20(8)& 1.72(15) & 8/8 & 3.9(6) \\
& 0.13(9) & 0.10(2) & 15/9 & {\bf 34(10)} & {\bf 0.26(9)}&   {\bf 7/8} & 0.02 & ? & PO & 0.14(3) & 1.69(12) & 62/64 &  3.3(3)  \\\vspace{0.07in}
& 0.17(5) & 0.16(2) & 31/9 & {\bf 29(4)} & {\bf 0.33(5)} & {\bf 6/8} & 4 E$-$4 & A & PO & 0.15(3) &  1.67(9) & 58/61 & 3.4(2)\\

18 & 0.20(12) & 0.10(2) & 17/12  & {\bf 25(8)} & {\bf 0.19(5)} & {\bf 15/11} & 0.24 & ? & PO & 0.12{2} & 1.74(7) & 104/96 & 8.9(5)\\
& 0.08(6) & 0.09(3) & 12/9 & {\bf 26(10)} & {\bf 0.20(7)} & {\bf 10/8} & 0.20 & ? & PO & 0.11(3) & 1.69(12) & 29/29 & 8.8(8)\\
 &  0.09(4) & 0.12(2) & 24/12 & {\bf 30(7)} & {\bf 0.31(8) }& {\bf 15/11} & 0.03 & ?& PO & 0/114(9) & 1.74(6) & 189/160 & 8.2(3) \\\vspace{0.07in}
& 0.10(2) & {\bf 0.11(2)} & {\bf 17/12} & 32(6) & 0.23(5) & 19/11 & NA & ? & PO & 0.12(2) & 1.79(5) & 146/173 & 8.7(3) \\

19 &0.20(12) &  {\bf 0.07(2) } & {\bf 13/12} & 32(14) & 0.14(8) & 15/11 & NA & B & PO & 0.15{7} & 2.9(4) & 23/37 & 3.3(4)\\
 & 0.1(4) & {\bf 0.11(4)} & {\bf 15/9} & 43.6(12) & 0.34(13) & 16/8 &  NA & ? & PO & 0.1 f & 2.3(2) & 10/9 & 2.9(6) \\
 & 0.20(9) & 0.13(2) & 9/9 & {\bf 22(8) } & {\bf 0.22(5)}& {\bf 6/8} & 0.07 & B& PO & 0.20(5) & 3.0(3) & 130/105 &5.1(5) \\\vspace{0.07in}
&0.11(4) & 0.16(2) & 40/9 & {\bf 22(3)} & {\bf 0.31(4)} & {\bf 9/8} & 6 E$-$4 & A &PO &  0.23(2)  & 2.47(9) & 182/115 & 9.7(5)\\

20 & 0.07(7) & {\bf 0.01(3)} & {\bf 6/9}  & 44(2) & 0.42(7) & 6/8 & 0.63 & $-$ & PO & 0.14(2) & 1.86(9) & 85/86 & 7.8(5)\\
& 0.10(7) & 0.14(3) & 8/9 & {\bf 35(13)} & {\bf 0.4(2)} & {\bf 7/8} & 0.23 & ? & PO & 0.13(4) & 1.83(14) & 47/33 &8.0(6)\\
 & 0.09(5) & 0.10(2) & 8/9 & {\bf 35(10)} & {\bf 0.26(9)} & {\bf 6/8} & 0.15 & $-$& PO & 0.14(2) & 1.84(7) & 142/127 & 7.1(3)\\\vspace{0.07in}
& 0.08(3) & 0.11(2) & 20/9 & {\bf 38(5)} & {\bf 0.30(6)} & {\bf 9/8} & 0.02 & ? & PO & 0.16(2) & 1.80(5) & 223/201 & 9.1(3)  \\

\noalign{\smallskip}
\hline
\noalign{\smallskip}
\end{tabular}

\end{table*}

\setcounter{table}{1}
 \begin{table*}[!t]   
\renewcommand{\baselinestretch}{.5}
\renewcommand{\tabcolsep}{1mm}
\caption{ continued }
\begin{tabular}{llllllllllllllll}
\noalign{\smallskip}
\hline
\noalign{\smallskip}
& & \multicolumn{2}{c}{Power Law Fit} & \multicolumn{3}{c}{Broken Power Law Fit}& & &\multicolumn{5}{c}{SED FIT} \\
S & { $F_{\rm var}$} & $\gamma$ & $\chi^2_{\nu}$ & $\nu_{\rm c}$ & $\beta$ & $\chi^2_\nu$& $F$  & Type & Spec& $N_{\rm H}$ & $\Gamma$/kT & $\chi^2_\nu$ &   L\\
\noalign{\smallskip\hrule\smallskip}
21 & 0.27(17) & 0.11(2) & 21/9  & {\bf 19(3)} & {\bf 0.22(3)} & {\bf 7/8} & 6 E$-$3 & ? & PO & 0.1 f & 1.8(2) & 29/18 & 1.0(2)\\
 & 0.21(17) & {\bf 0.06(4)}& {\bf 5/9} & 20(20) & 0.09(8) & 5/8 & NA & $-$ & PO & 0.6(2) & 2.6(4) & 9/5 & 3.2(6) \\
& 0.14(8) & 0.10(2) & 36/12 & {\bf 48(7)} & {\bf 0.58(17)} & {\bf 10/11} & 2 E$-$4 & A$^{\dagger}$ & PO & 0.28(4) & 2.0(13) & 62/71 &  3.4(3)\\\vspace{0.07in}
& 0.20(5) & 0.15(2) & 47/9 & {\bf 37(4) } & {\bf 0.44(6) } & {\bf 11/8} & 2 E$-$3 & A$^{\ast}$ & PO & 0.28(4) & 2.15(9) & 45/50 & 2.2(3)\\

22 & 0.13(5)&  {\bf 0.07(2)} & {\bf 10/9}  & 40(20) & 0.19(15) & 12/8 & NA & B & 2C & 0.1 f & 2.50(9) & 77/73 & 6.9(3)\\
 & 0.16(14) & 0.16(4) & 11/9 & {\bf 46(10)} & {\bf 0.7(3)} & {\bf 4/8} & 7 E$-$3 & ? & PO & 0.1 f & 1.4(2) & 9/15 & 5.1(14) \\
 & 0.10(5) & 0.11(2) & 38/12 & {\bf 26(6) } & {\bf 0.27(6)} & {\bf 24/11} & 0.03 & ? & 2C & 0.1 f & 2.42(8) & 127/128 & 6.6(12)\\\vspace{0.07in}
&0.16(3) & 0.13(2) & 53/9  & {\bf 29(4) } & {\bf 0.33(5)} & {\bf 7/8} & 3 E$-$5 & A$^{\dagger}$ & 2C & 0.1 f & 2.40(6) & 139/114  & 6.8(5)\\

23 & 0.12(8) & 0.05(2) & 19/12  & {\bf 36.8(3)} &{\bf 0.20(6)} & {\bf 12/11}  & 0.03 & ? & PO & 0.06(3) & 2.14(0.15) &43/54 &  2.8(3)\\
 & 10(10) & 0.22(3) & 38/8 & {\bf 45(6)} & {\bf 1.0(3)} & {\bf 9/8} & 9 E$-$4 & A & $\dots$& $\dots$& $\dots$& $\dots$ & $\dots$ \\
& 5(5) & 0.01(2) & 14/9 & {\bf 33(8)} & {\bf 0.24(5) } & {\bf 4/8} & 2 E$-$3 & ?&  $\dots$& $\dots$& $\dots$  \\\vspace{0.07in}
& 0.9(16) & 0.04(2) & 20/12 & {\bf 17(8)} & {\bf 0.09(3)} & {\bf 16/11} & 0.12 & ?  & $\dots$& $\dots$& $\dots$ & $\dots$ & $\dots$\\

24 & 0.04(106) & {\bf 0.02(2)} & {\bf 7/9}  & 60(90) & 0.1(3) & 7/8& NA & $-$ & PO & 0.20(11) & 1.7(3) & 11/17 & 1.4(3)\\
 & 0.31(19) & 0.06(3) & 13/12 & {\bf 30(20)} & {\bf 0.18(11)} & {\bf 11/11} & 0.15 & $-$& PO & 0.15(14) & 1.5(4) & 18/15 &  1.3(3)\\
 & 0.31(15) & 0.04(2) & 9/12 & {\bf 36(17) } & {\bf 0.12(8)} & {\bf 9/11} & 0.30 & $-$ & PO & 0.32(8) & 1.9(3) & 46/38 & 1.5(2) \\\vspace{0.07in}
&0.27(9) & 0.18(2) & 57/9 & {\bf 31(3)} & {\bf 0.45(5)} & {\bf 10/8} & 2 E$-$4 & A & PO & 0.20(5) & 1.69(15) & 23/20 & 1.11(2)  \\

25 & 0.02(4) & 0.04(2) & 10/9 & {\bf 52(12)} & {\bf 0.25(16)} &{\bf 6/8} & 0.06 &  $-$ & 2C & 0.152(5) & 2.11(5) & 423/416 & 58(3) \\
& 0.02(3) & {\bf 0.06(3)} & {\bf 10/12} & 40(60) & 0.08(17) & 12/11 & NA & $-$ & 2C & 0.134(11) & 2.02(6)  & 189/204 & 63(4) \\
& 0.038(12) & 0.19(2) & 18/9 & {\bf 0.02(4)} & {\bf 0.32(5)} & {\bf 7/8} & 9 E$-$3 & ?& 2C & 0.162(12)& 2.23(3) & 715/666 & 57(3) \\\vspace{0.07in}
& 0.093(9) & 0.19(2) & 27/9 & {\bf 34(4) }& {\bf 0.94(7)} & {\bf 21/8} & 0.15 & B & 2C & 0.13(9) & 2.13(8) & 598/557 & 51(2)\\

26 & 0.1(4) & -0.02(2) & 11/12  & {\bf 57.8(11)} & {\bf 0.12(13)} & {\bf 10/11} & 0.58 & $-$& PO  & 0.19(9) & 2.7(3) & 36/33 & 1.5(3)\\
& 0.3(8) & 0.15(4) & 9/9 & {\bf 44(2)} & {\bf 0.40(12)} & {\bf 9/8} & 0.52 & ? & PO&  0.15 f & 3.1(7) &25/23 & 0.9(5) \\
 & 0.3(13) & 0.05(2) & 25/12 & {\bf 25(9)} & {\bf 0.18(6)} & {\bf 14/11} & 0.08 & ? & PO & 0.07(7) & 2.5(7) & 41/51 & 0.43(10)\\\vspace{0.07in}
& 0.26(13) & 0.08(2) & 23/12 & {\bf 35(7)} & {\bf0.24(6) } & {\bf 6/11} & 2 E$-$4 & A$^{\dagger}$ & PO & 0.23(6) & 2.9(3) & 62/64 &1.7(2) \\

27 & 0.9(5) & 0.15(2) & 14/9  & {\bf 36(6)} & {\bf 0.38(9)} & {\bf 8/8} & 0.04 & ?  & PO & 0.3 f & 1.0(4) & 42/30  &0.7(4)\\
 & 2(10) & 0.04(3) & 15/9 & {\bf 57.1(8)} & {\bf 0.3(2) } & {\bf 12/8} & 0.19 &? &PO &  0.1 f & 1.3(13) & 24/14 & 0.4(3)\\
& 1.3(7) & 0.20(2) & 40/12 & {\bf 30(4)} & {\bf 0.45(6)} & {\bf 19/11} & 5 E$-$3 & A$^{\ast}$ & PO & 0.6(4) & 1.5(5)&  40/40 & 0.7(2) \\\vspace{0.07in}
& 0.42(11) & 0.15(2) & 36/9 & {\bf 20(4)} & {\bf 0.26(4)} & {\bf 7/8} & 5 E$-$4 & A & PO & 0.40(11) & 1.80(15)& 31/38 & 1.4(2)\\

28 & 1.1(4) & 0.18(2) & 34/9 &  {\bf 35(5)} & {\bf 0.54(10)} & {\bf 6/8} & 4 E$-$4 & A$^{\dagger}$ & PO & 0.2 f & 1.3(5) & 7/7 & 0.4(3)\\
& 0.4(6) & {\bf 0.07(4) } & {\bf 22/9} & 57.1(2) & 0.1(2) & 24/8 & NA & ? & PO & 0.2 f & 1.7(7) & 1/3 &0.4(2) \\
& 0.9(3) & 0.21(2) & 59/12 & {\bf 27(3)} &{\bf 0.55(7)} &{\bf 5/8} & 3 E$-$4 & A & PO & 0.20(16) & 2.0(4) & 27/24 & 0.4(3) \\\vspace{0.07in}
& 0.79(17) & 0.21(2) & 77/9 & {\bf 34(3) } & {\bf 0.53(6) } & {\bf 18/8} & 9 E$-$4 & A & PO & 0.2(2) & 1.5(3) & 15/8 & 0.52(10)\\

29 & 0.2(2) & 0.001(2) & 10/12 &  {\bf 46.5(9)} &{\bf  0.06(8)} & {\bf 10/11} & 0.42 & $-$ &  NSA  & 0.1 f & 0.068(10) & 32/25 &  0.6(5)\\
& 0.35(17)&   0.15(3) & 24/12 & {\bf 29(3) } & {\bf 0.43(7)} & {\bf 10/11} & 2 E$-$3 & ? & NSA & 0.1 f & 0.09(4) & 6/6 & 0.8(3)\\
& 0.32(16) & 0.11(2) & 34/12 & {\bf 46(5)} & {\bf 0.50(11)} & {\bf 22/11} & 0.03 & ? & NSA & 0.1f & 0.065(7) & 55/48 &  0.6(4)\\\vspace{0.07in}
& 0.61(16) & 0.21(2) & 73/9 & {\bf 38(3)} &{\bf 0.60(7)} & {\bf 18/8} & 1 E$-$3 & A$^{\ast}$ & NSA & 0.1 f & 0.068(12) & 39/28 & 0.3(2) \\

30 & 0.09(5) & 0.12(2) & 15/9  & {\bf 26(13)} & {\bf 0.23(4)} & {\bf 10/8} & 0.09 & ? & PO & 0.1 f & 2.13(9) & 52/41 & 2.7(3) \\
 & 0.10(14) & 0.06(3) & 10/12 & 17(13) & 0.12(5) & 8/11 & 0.13 & $-$ & PO & 0.1 f & 2.22(14) & 18/12 &  2.8(4)\\
& 0.08(11) & {\bf 0.01(2) }& {\bf 8/9} & 70(14) & 0.3(4) & 8/8 & NA & $-$ &PO & 0.13(3) & 2.43(16) & 79/81 & 3.0(2)\\\vspace{0.07in}
& 0.12(4) & 0.14(2) & 21/9  & {\bf 14(3)} & {\bf 0.20(2)} & {\bf 9/8} & 9 E$-$3 & ? & 2C & 0.06(3) & 1.88(10) & 133/94 & 3.3(2)\\

\noalign{\smallskip}
\hline
\noalign{\smallskip}
\end{tabular}

\end{table*}

\setcounter{table}{1}
 \begin{table*}[!t]   
\renewcommand{\baselinestretch}{.5}
\renewcommand{\tabcolsep}{1mm}
\caption{  continued}
\begin{tabular}{llllllllllllllll}
\noalign{\smallskip}
\hline
\noalign{\smallskip}
& & \multicolumn{2}{c}{Power Law Fit} & \multicolumn{3}{c}{Broken Power Law Fit}& & &\multicolumn{5}{c}{SED FIT} \\
S & { $F_{\rm var}$} & $\gamma$ & $\chi^2_{\nu}$ & $\nu_{\rm c}$ & $\beta$ & $\chi^2_\nu$& $F$  & Type & Spec& $N_{\rm H}$ & $\Gamma$/kT & $\chi^2_\nu$ &   L\\
\noalign{\smallskip\hrule\smallskip}
31 & 0.09(5) & 0.140(2) & 24/9  & {\bf 48(7) } & {\bf 0.51(15)} & {\bf 19/8} & 0.20 & ? & 2C & 0.25(8) & 2.1(2) & 86/90 & 6.9(7)\\
 & 0.06(5) & 0.10(3) & 26/9 &{\bf 43(8) } & {\bf 0.52(17)} & {\bf 12/8} & 0.02 & ?& PO & 0.15(3) & 2.0(10) & 56/55 & 15.7(11) \\
& 0.06(2) & 0.11(2) & 26/12 & {\bf 27(6)} & {\bf 0.28(6)} & {\bf 11/11} & 2 E$-$3 & ? & 2C & 0.206(9)& 2.54(6)  & 446/391 & 23.9(10) \\\vspace{0.07in}
& 0.07(2) & 0.15(2) & 39/9 & {\bf 28(4)} & {\bf 0.34(5)} & {\bf 6/8} & 2 E$-$4 & A & 2C & 0.20(2) & 2.40(5) & 353/314 &16.5(4) \\

32 & 0.34(12) & 0.16(2) & 24/9  & {\bf 40(60)} & {\bf 0.44(10)} & {\bf 14/8} & 0.05 & ? &PO &  0.43(0.16) & 1.7(3) & 40/34 & 3.6(5)\\
 & 0.18(10) & 0.13(3) & 14/12 &{\bf 31(13)} &{\bf 0.29(15)} & {\bf 13/11 }& 0.63 & ? & PO  & 0.36(16) & 1.6(2) & 11/10 & 6.0(9)\\
&0.11(6) & 0.11(2) & 8/9 & {\bf 17(2)} & {\bf 0.18(3) } & {\bf 4/8} & 0.02 &  B& PO & 0.4(6)&  1.84(13) &148/149 & 9.8(7)   \\\vspace{0.07in}
&0.04(12) & {\bf 0.07(2)} & {\bf 11/12} & 19(6) & 1.00(3) & 36/11 & NA & $-$ & PO & 0.42(3) & 1.95(8) & 120/106 & 9.4(5)\\

33 & 0.02(13) & {\bf 0.02(2) }& {\bf 16/12 } & 36.8(5) & 0.03(6) & 18/11 & NA & $-$ & PO & 0.09(2) & 1.87(7) & 146/153 & 11.8(7)\\
& 0.07(9) & 0.1(3) & 12/9 & {\bf 14(9)} & {\bf 0.23(5)} & {\bf 11/8} & 0.53 & ? & PO& 0.27(5) & 2.04(17)& 43/41 &   8.4(8)\\
 & 0.08(4) &{\bf  0.16(2)} & {\bf 8/9} & 18(6) & 0.23(5) & 9/8 & NA & B &PO&  0.09(2) & 1.95(6) & 213/182 & 10.7(5)\\\vspace{0.07in}
& 0.10(2) & 0.23(2) & 48/9 & {\bf 31(3) } & {\bf 0.49(5)} & {\bf 12/8 } & 1 E$-$3 & A$^{\ast}$ & PO & 0.14(2) & 1.97(4) & 350/309 &10.7(3) \\

34 & 0.05(6) & 0.02(2) & 5/9  & {\bf 50(30) } & {\bf 0.10(16)} & {\bf 4/8} & 0.47 & $-$ &PO & 0.20(2) &  2.03(7) & 112/124 & 15.3(6)\\
 & 0.04(11) & 0.13(4) & 18/9 & {\bf 44(10)} & {\bf 0.61(13)} & {\bf 7/8} & 6 E$-$3 & ? & PO &  0.18(4) & 1.96(12) & 27/35 & 10.4(8)\\
& 0.06(4) & {\bf 0.14(2) } & {\bf 6/9} & 27(7) & 0.26(7) & 7/8 & NA & B & PO & 0.16(2) & 1.78(5) & 172/165 & 13.6(5)\\\vspace{0.07in}
& 0.11(3) & 0.09(2)& 19/9 & {\bf 39(6)} & {\bf 0.27(7)} & {\bf 5/8} & 2 E$-$3 & ? & PO & 0.22(2) & 2.20(5) & 132/132 & 7.0(3)\\

35 & 0.08(5) & 0.16(2) & 14/9 & {\bf 12(4)} & {\bf 0.23(4)} & {\bf 11/8} & 0.24 & ? & NSA & 0.18(2) & 0.017(1) & 102/80 & 38(3)\\
 & 0.12(8) & 0.12(3) & 23/9 & {\bf 33(11) } & {\bf 0.40(18)} & {\bf 10/8} & 0.02 & ?& NSA & 0.17(4) & 0.017(1) & 34/28 & 26(6)\\
 & 0.10(6) & {\bf 0.07(2)} & {\bf 17/9} & 3(50) & 0.02(2) & 17/8 & NA & $-$ & NSA & 0.16(2)  & 0.0164(4) & 86/76 & 32(2) \\\vspace{0.07in}
&0.08(4)&  0.07(2) & 25/12 & {\bf 27(2)} & {\bf 0.14(3)} & {\bf 18/11} & 0.06 & ? & NSA$^{e}$ & 0.16(2) &0.018(2) & 60/77 & 24(7)  \\

36 &  $\dots$ &  $\dots$ &  $\dots$  &  $\dots$ &  $\dots$ &  $\dots$ &  $\dots$ &  $\dots$ &  $\dots$ &  $\dots$ &  $\dots$ &  $\dots$ &  $\dots$  \\
&  $\dots$ &  $\dots$ &  $\dots$  &  $\dots$ &  $\dots$ &  $\dots$ &  $\dots$ &  $\dots$ &  $\dots$ &  $\dots$ &  $\dots$ &  $\dots$ &  $\dots$ \\
&  $\dots$ &  $\dots$ &  $\dots$  &  $\dots$ &  $\dots$ &  $\dots$ &  $\dots$ &  $\dots$ &  $\dots$ &  $\dots$ &  $\dots$ &  $\dots$ &  $\dots$ \\\vspace{0.07in}
& 0.3(6) &{\bf  0.11(2) }&{\bf 13/9} &33(2) & 0.25(3) & 13/8 & NA & ?$^{f}$ & PO & 0.27(7) & 1.5(2) & 46/29 & 4.1(5)\\

37 &  0.06(6) & 0.06(2) & 25/9  & {\bf 50(12)} & {\bf 0.29(15)} & {\bf 20/8} & 0.23 & ? & PO & 0.15(2) & 1.88(8) & 128/119 & 9.3(6)\\
& 0.02(19)& 0.02(3) & 16/12& {\bf 38(2)} & {\bf 0.10(10)} & {\bf 16/11} & 0.43 & $-$ & PO & 0.11(3) & 1.84(12) & 48/35 & 8.9(7) \\
& 0.12(4) & 0.131(14) & 34/12 & {\bf 34(4)} & {\bf 0.34(6)} & {\bf 18/11} & 0.01 & A$^{\ast}$ & PO & 0.17(2) & 1.89(6) &233/195 &  8.6(3) \\\vspace{0.07in}
&0.08(3) &  0.21(2) & 51/9 & {\bf 35(3)} & {\bf 0.50(6)} & {\bf 7/8} & 2 E$-$4 &  A & PO & 0.15(2) & 1.84(5) &204/186 &  8.7(3)\\

38 & 0.19(6) & 0.22(2) & 19/9  & {\bf 27(4)} & {\bf 0.44(7)} & {\bf 9/8} & 0.02 & ? & PO & 0.15(4) & 1.72(17) & 109/85 & 4.8(5)\\
&  0.2(2) & 0.08(3) & 31/12 & {\bf 44(8)} & {\bf 0.54(19)} & {\bf 16/11} & 8 E$-$3 & ? &PO & 0.13(13) & 2.2(5) & 4/5 & 2.1(5) \\
&0.22(18) & 0.10(2) & 30/12 & {\bf 44(5)} & {\bf 0.44(11)} & {\bf 14/11} & 0.06 & ?& PO & 0.26(9) & 3.2(7) & 67/83 & 2.7(4) \\\vspace{0.07in}
& 0.24(5) & 0.18(2) & 50/9 & {\bf 35(3)} & {\bf 0.47(6)} & {\bf 7/8} & 2 E$-$4 & A & PO & 0.17(4) & 2.08(15) & 143/87 & 2.8(3)\\

39 & 0.2(3)& 0.14(2) & 29/12  &  {\bf 39(6)} & {\bf 0.43(11)} & {\bf 20/11} & 0.04&?& PO & 0.1 f & 1.7(4) & 12/25&  0.6(3)\\
 & 0.09(51) & 0.13(3) & 18/9 & {\bf 34(10)} & {\bf 0.41(17)} &{\bf 8/8} & 0.02 & ? & PO & 0.1 f & 1.9(3) & 7/7 & 1.2(5)\\
& 0.1(6) & {\bf 0.07(2)} & {\bf 12/9} & 32(2) & 0.15(5) & 13/8 & NA & $-$ & PO & 0.10(6) & 1.80(16) & 24/49 & 1.2(2)  \\\vspace{0.07in}
& 0.1(3) & 0.12(2) & 38/12 & {\bf 31(5)} & {\bf 0.32(6)} & {\bf 6/11} & 9E$-$5 & A$^{\ast}$ & PO & 0.08(4) & 1.7(2) & 44/46 & 0.97(14)\\

40 & 0.15(6) & 0.12(2) & 18/9  & {\bf 37(7)} & {\bf 0.37(10)} & {\bf 8/8} & 0.01 & ? & PO & 0.15(3) & 2.02(13) & 60/45 & 4.2(3)\\
&0.16(8) & 0.06(3) & 9/9 & {\bf 51(16)} & {\bf 0.3(2)} & {\bf 7/8} & 0.16 & ? & PO & 0.14(7) & 2.1(3) & 9/12 & 4.3(5)\\
&0.09(6) &  0.18(2) & 56/9 & {\bf 41(3)} & {\bf 0.60(6) } & {\bf 11/8} & 4 E$-$4 & A & PO & 0.14(2) & 1.88(9) & 93/99 & 4.9(3)\\\vspace{0.07in}
& 0.11(5) & {\bf 0.06(2) } & {\bf 14/12} & 6.8(3) & 0.05(2) & 15/11 & NA & $-$ & PO & 0.15(2) & 2.09(8) & 85/92 & 4.3(2) \\

\noalign{\smallskip}
\hline
\noalign{\smallskip}
\end{tabular}
\\$^{e}$ Additional emission feature required, well fitted by 0.541(6) keV Gaussian with $\sigma$ = 0.036(8)\\
$^{f}$ Only PN data were available, not MOS
\end{table*}

\setcounter{table}{1}
 \begin{table*}[!t]   
\renewcommand{\baselinestretch}{.5}
\renewcommand{\tabcolsep}{1mm}
\caption{ continued. }
\begin{tabular}{llllllllllllllll}
\noalign{\smallskip}
\hline
\noalign{\smallskip}
& & \multicolumn{2}{c}{Power Law Fit} & \multicolumn{3}{c}{Broken Power Law Fit}& & &\multicolumn{5}{c}{SED FIT} \\
S & { $F_{\rm var}$} & $\gamma$ & $\chi^2_{\nu}$ & $\nu_{\rm c}$ & $\beta$ & $\chi^2_\nu$& $F$  & Type & Spec& $N_{\rm H}$ & $\Gamma$/kT & $\chi^2_\nu$ &   L\\
\noalign{\smallskip\hrule\smallskip}
41 & 0.35(18) & 0.09(2) & 13/12  & {\bf 43(8)} & {\bf 0.33(11) } & {\bf 11/11} & 0.18 &  ?& PO &0.06(6) &1.7(3) & 27/20 & 2.0(3)\\
& 0.2(5) & 0.10(4) & 5/9 &{\bf 37(16) } & {\bf 0.26(15)} &{\bf 3/8} & 0.06 & ? & PO & 0.1 f & 2.1(5) & 5/4 & 0.8(3) \\
& 0.38(18) & 0.13(2) & 22/9 & {\bf 42(5)} & {\bf 0.39(9)} & {\bf 17/8} & 0.17 & ? & PO & 0.1 f & 2.3(4) & 23/22 & 0.7(3)\\\vspace{0.07in}
& 0.38(11) & 0.18(2) & 43/9 & {\bf 26(3) } & {\bf 0.36(4)} & {\bf 6/8} & 9 E$-$5 & A & PO & 0.09(5) & 1.71(15) & 51/43 & 0.97(14)\\

42 & 0.13(6) & 0.09(2) & 25/9  & {\bf 30(7)} & {\bf 0.28(7)} & {\bf 9/8}  & 5 E$-$3 & ? & PO & 0.18(3) & 2.18(14) & 53/52 & 4.4(3)\\
& 0.10(11) & 0.05(4) & 18/9 & {\bf 44(2) } & {\bf 0.29(12)} & {\bf 12/8} &  0.10 & ?& PO & 0.20 & 2.02(13) & 41/28 & 5.0(7)\\
& 0.04(17) & 0.08(2) & 8/9 & {\bf 60(40)} & {\bf 0.2(3)} & {\bf 7/8} & 0.3 & $-$ & PO & 0.12(2) & 1.87(9) & 121/111 & 4.9(3)\\\vspace{0.07in}
& 0.12(3) & 0.17(2) & 34/9 & {\bf 30(4)} & {\bf 0.36(5)} & {\bf 7/8} & 6 E$-$4 & A$^{\dagger}$ & PO & 0.16(2) & 2.02(7) & 131/110 & 5.2(2)\\

43 & 0.09(10) & 0.06(2) & 7/9 & {\bf 18(13) } & {\bf 0.1(4) } & {\bf 5/8} & 0.13 & $-$ & PO & 0.13(3) & 1.56(10) & 58/60 & 4.8(4)\\
& 0.02(40)& 0.18(3) & 19/9 & {\bf 46(8)} & {\bf 0.81(3)} &{\bf 4/8} & 4 E$-$4 & A & PO & 0.17(8) & 1.7(2) & 12/18 & 5.1(6)\\
& 0.10(6) & 0.11(2) & 39/12 & {\bf 31(5)} & {\bf 0.37(7)} & {\bf 8/11} & 4 E$-$5 & A & PO & 0.13(2) & 1.58(8) & 114/126 & 5.5(3) \\\vspace{0.07in}
& 0.14(3) & 0.20(2) & 47/9 & {\bf 34(4)} & {\bf 0.49(2)} & {\bf 20/8} & 0.01 & ? & PO & 0.14(2) & 1.58(6) & 166/138 & 5.8(2)\\

44 &  $\dots$ &  $\dots$ &  $\dots$ &  $\dots$ &  $\dots$ &  $\dots$ &  $\dots$ &  $\dots$ &  $\dots$ &  $\dots$ &  $\dots$ &  $\dots$ &  $\dots$  \\
&  $\dots$ &  $\dots$ &  $\dots$ &  $\dots$ &  $\dots$ &  $\dots$ &  $\dots$ &  $\dots$ &  $\dots$ &  $\dots$ &  $\dots$ &  $\dots$ &  $\dots$\\
&  $\dots$ &  $\dots$ &  $\dots$ &  $\dots$ &  $\dots$ &  $\dots$ &  $\dots$ &  $\dots$ &  $\dots$ &  $\dots$ &  $\dots$ &  $\dots$ &  $\dots$\\\vspace{0.07in}
& 0.21(14) & {\bf 0.04(2)} & {\bf 11/9} & 14(8) & 0.05(2) & 12/8 & NA & $-^{f}$ & PO & 0.12(5) & 0.98(10) & 47/48 & 6.5(5)\\

45 & 1.8(5) & 0.20(2) & 34/9  & {\bf 45(5) } &{\bf 0.68(17) } & {\bf 12/8} & 5 E$-$3 & A$^{\ast}$ &PO &  0.9(9) & 0.9(10) & 18/12 & 0.5(3)\\
& 0.5(12) & 0.08(3) & 15/9 &{\bf 37(2) } & {\bf 0.24(9)} & {\bf 14/8} & 0.45 & ? & PO& 0.1 f & 1.3(9) & 10/6 & 0.4(3) \\
& 1.0(4) & {\bf 0.08(2) } & {\bf 4/9} & 9(7) & 0.10(4) & 5/8 & NA & B & PO & 0.6(6) & 1.5(9) & 16/22 & 0.5(2)\\\vspace{0.07in}
& 3.5(6) & 0.23(2) & 91/9 & {\bf 30(2)} & {\bf 0.53(4)} & {\bf 5/8} & 3 E$-$6 & A & PO & 3(2) & 3(2) & 2/6 & 0.4(2)\\

46 & 0.3(2) & {\bf 0.09(2)} & {\bf 4/12} & 50(5) & 0.10(2) & 4/11 & NA & B & PO & 0.1 f & 1.9(4) & 9/7 & 1.9(7)\\
& 0.5(5) & {\bf 0.15(4)} & {\bf 14/12} & 46(14) & 0.3(2) & 17/11 & NA & ?& PO & 0.1 & 2.2(5) & 4/5 & 0.6(3) \\
& 0.2(2) & 0.03(2) & 14/12 & {\bf 29(21)} & {\bf 0.08(6) } & {\bf 13/11} & 0.41 & ? & 2C & 0.1 f & 2.5(2) & 35/24 &2.32(10)\\\vspace{0.07in}
& 0.55(10) & 0.22(2) & 41/9 &  {\bf 25(3) } & {\bf 0.39(4)} & {\bf 10/8}& 1 E$-$3 & A$^{\ast}$ & PO & 0.1 f & 1.83(17) & 17/13 & 0.59(10)\\

47 & 0.13(6) & 0.14(2) & 15/9 &{\bf 37(7)} &{\bf 0.34(10)} & {\bf 11/8} & 0.14 & ? & PO & 0.10(3) & 2.28(15) & 65/54 & 3.5(3)\\
& 0.09(12) & {\bf 0.07(4) } & {\bf 12/9} & 19(20) & 0.05(10) & 13/8 & NA & $-$ & PO & 0.09(5) & 1.9(2) & 20/18 & 4.6(6)\\
 & 0.13(6) & 0.15(2) & 42/9 & {\bf 41(4)} & {\bf 0.59(9)} & {\bf 5/8} & 6 E$-$5 & A &PO &  0.12(2) & 2.01(11) & 110/97 & 3.4(2)\\\vspace{0.07in}
& 0.11(2) & 0.21(2) & 62/ 9 & {\bf 21(3) } & {\bf 0.38(4)}& {\bf 11/8} & 3 E$-$4 & A & PO & 0.15(2) & 1.89(4) &265/231 & 11.9(3)\\

48 & 0.19(9)& 0.09(2) & 22/9 & {\bf 32(7)} & {\bf 0.28(7)} &{\bf 8/8}& 7 E$-$3 & ? &  PO & 0.09(4) & 2.2(2) & 29/24 & 1.9(3) \\
& 0.5(2) & 0.13(3) & 23/9 & {\bf 48(8)} &{\bf 0.7(3)} &{\bf 10/8} & 0.01 & ? & PO & 0.1 f & 2.1(3) & 12/10 & 1.1(3) \\
& 0.18(10) & 0.12(2) & 63/9 & {\bf 42(2) } & {\bf 0.51(5) } & {\bf 21/8} & 4 E$-$3 & ? & PO & 0.11(4)& 2.3(3) & 53/50 & 1.6(2)\\\vspace{0.07in}
& 0.36(9) & 0.19(2) & 42/9 & {\bf 22(3)} & {\bf 0.34(4)} & {\bf 6/8} & 1 E$-$4 & A & PO & 0.14(4) & 2.2(2) & 41/35 & 1.6(2)\\

49 & 0.2(2) & 0.08(2)& 16/9 & {\bf 35(10)}&{\bf 0.24(9) } &{\bf 8/8} & 0.03 & ? & $\dots$& $\dots$& $\dots$\\
& 0.17(9)& 0.15(4) & 12/8 & {\bf 48(10)} & {\bf 0.6(3)} & {\bf 5/8} & 0.01 & ? &PO & 0.12(8) & 1.9(3) & 11/13 & 3.9(6) \\
& 0.20(11) & {\bf 0.15(2)} & {\bf 17/9} & 41(7) & 0.34(9) & 26/8 & NA & ?  & $\dots$& $\dots$& $\dots$\\\vspace{0.07in}
& 0.17(5) & 0.19(2) & 63/9 & {\bf 37(3) } & {\bf 0.53(6)} & {\bf 12/8} & 4 E$-$4 & A & PO & 0.13(4) & 1.87(13) & 52/37 & 2.4(2)\\

50 & 0.1(10) & 0.07(2) & 33/12 & {\bf 27(7)} & {\bf 0.27(6)} & {\bf 13/11} & 1 E$-$3 & A$^{\ast}$ & PO & 0.1 f & 2.2(4) & 8/14 &  0.9(3)\\
 & 0.1(9) & 0.18(3) & 14/9 & {\bf 20(8) } & {\bf 0.31(9) } & 11/8 & 0.21 & ? & PO & 0.3(2) & 3.1(9) & 3/7 & 1.2(3)\\
&  $\dots$ &  $\dots$ &  $\dots$ &  $\dots$ &  $\dots$ &  $\dots$ &  $\dots$ &  $\dots$ &  $\dots$ &  $\dots$ &  $\dots$ &  $\dots$ &  $\dots$\\\vspace{0.07in}

& 0.33(10) &{\bf  0.10(2)} &{\bf 5/9} & 3(40) & 0.1(2) & 5/8 & NA & B & PO & 0.07(4) & 1.78(14) &39/40 & 1.07(10)\\
\noalign{\smallskip}
\hline
\noalign{\smallskip}
\end{tabular}

\end{table*}

\setcounter{table}{1}
 \begin{table*}[!t]   
\renewcommand{\baselinestretch}{.5}
\renewcommand{\tabcolsep}{1mm}
\caption{ continued. }
\begin{tabular}{llllllllllllllll}
\noalign{\smallskip}
\hline
\noalign{\smallskip}
& & \multicolumn{2}{c}{Power Law Fit} & \multicolumn{3}{c}{Broken Power Law Fit}& & &\multicolumn{5}{c}{SED FIT} \\
S & { $F_{\rm var}$} & $\gamma$ & $\chi^2_{\nu}$ & $\nu_{\rm c}$ & $\beta$ & $\chi^2_\nu$& $F$  & Type & Spec& $N_{\rm H}$ & $\Gamma$/kT & $\chi^2_\nu$ &   L\\
\noalign{\smallskip\hrule\smallskip}
51 & 1.5(8) & 0.08(2) & 13/12 & {\bf 45(10)} & {\bf 0.28(11)} & {\bf 11/11} & 0.22 & ? & NSA  & 0.1 f & 0.08(6) & 29/25 & 0.08(8)\\
& 3(2) & 0.22(3) & 11/12 & {\bf 24(6)} & {\bf 0.47(12)} & {\bf 4/11} & 6 E$-$4 & A$^{\dagger}$ & NSA & 0.1 f & 0.08(5) & 3/5 & 0.8(8)\\
& 1.4(12) & 0.14(2) & 23/9  & {\bf 48(6)} & {\bf 0.58(15)} &{\bf 4/8} & 2 E$-$4 & A & PO & 0.1 f & 4(4) & 30/23 & 0.4(4) \\\vspace{0.07in}
& 0.0.09(3) & 0.12(2) & 19/9 & {\bf 37(6)} & {\bf 0.30 (7) } & {\bf 10/8} & 0.03 & ? & 2C & 0.08(2) & 2.30(11)  & 150/147 & 5.5(3) \\

52 & 0.49(12) & 0.19(2) & 32/9 & {\bf 30(4)} & {\bf 0.47(8)} & {\bf 13/8} & 0.01 & A$^{\ast}$ & PO & 0.1 f & 2.2(6) & 18/12 & 0.7(3) \\
 & 0.4(3) & 0.07(3) & 11/12 & {\bf 0.56(12)} & {\bf 0.5(3)} & {\bf 10/11} & 0.73 & $-$ & PO & 0.1 f & 0.9(4) & 6/6 &1.2(6) \\
& 0.41(13) & 0.10(2) & 8/12 & {\bf 16(7)} & {\bf 0.15(4)}  & {\bf 7/11} & 0.29 & B & PO & 0.1 f & 1.2(2) & 47/31 & 1.1(3)\\\vspace{0.07in}
& 0.45(10) & 0.15(2)& 27/9  & {\bf 23(4) } & {\bf 0.25(4)} & {\bf 10/8} & 6 E$-$3 & ? & PO & 0.1 f & 1.08(15) & 29/20 & 0.93(17) \\

53 & 1.4(8) & 0.06(2)& 9/9 & {\bf 46(15)} & {\bf 0.23(14)}  & {\bf 6/8} & 0.09 & ? &  PO & 0.1 f & 3.5(8) & 21/19 & 0.21(14)\\
& 11(14)& 0.15(3) & 33/9 & {\bf 39(7)} & {\bf 0.61(17)} &{\bf 8/8} & 9 E$-$4 & A$^{\dagger}$ & NSA & 0.1 f  & 0.05(5) & 12/8 & 0.2(2) \\
& 2.0(14) & 0.10(2) & 10/12 & {\bf 27(2)} & {\bf 0.18(4)} & {\bf 6/8} & 0.01 & ?& PO & 0.1 f & 3.6(13) & 19/19 & 0.16(12)\\\vspace{0.07in}
& 0.18(4) & 0.21(2) & 35/9 & {\bf 24(3)} & {\bf 0.37(4)} & {\bf 11/8} & 3 E$-$3 & A$^{\ast}$ & PO & 0.17(3) & 1.83(8) & 88/85 & 4.2(2)\\

54 & 0.08(3) & 0.15(2) & 6/9 & {\bf 90(50)} & {\bf 0.18(3) } & {\bf 3/8} & 0.03 & ? &  PO &  0.080(14) & 1.58(6) & 118/129 & 13.3(6)\\
& 0.01(29) & 0.09(4) & 15/9 &{\bf 28(14)} & {\bf 0.24(12)} &{\bf 12/8} & 0.14 & ? & PO & 0.11(3)& 1.71(11) & 37/43  & 13.2(9) \\
& 0.07(4) & 0.03(2) & 13/9 & {\bf 12(2) } & {\bf 0.05(3) } & {\bf 12/8} & 0.47 & $-$& PO & 0.10(2) & 1.79(4) & 204/203 & 11.4(4) \\\vspace{0.07in}
& 0.08(12) & 0.19(2) & 46/9 & {\bf 38(3)} & {\bf 0.52(6)} & {\bf 18/8} & 7 E$-$3 & ? & PO & 0.11(2) & 1.72(2) & 216/200 & 10.3(3)\\

55 & 0.22(5) & {\bf 0.11(2)} & {\bf 9/9} & 14(2) & 0.13(3) & 10/8 & NA & B &  PO & 0.10(5) & 0.64(5) & 76/62 & 14.9(11)\\
& 0.12(7) & 0.16(3) & 20/12 & {\bf  37(2)} & {\bf 0.49(8)} & {\bf 10/11} & 6 E$-$3 & ? & 2C & 0.2(2) & 2.1(9) & 22/23 & 21(6) \\
& 0.07(6) & 0.10(2) & 6/9 & {\bf 27(8)} & {\bf 0.21(7)} & {\bf 2/8} & 0.01 & ? & 2C & 0.1 f & 2.4(3) & 156/169 & 22(3)\\\vspace{0.07in}
& 0.45(4) & 0.17(2) & 34/12 &{\bf  22(4)} & {\bf 0.29(4)} & {\bf 21/11} & 0.02 & ? & PO & 0.05(4) & 0.52(7) & 89/72 & 12.1(8) \\

56 & 0.12(11) & 0.05(2) & 7/9 & {\bf 33(16)} &{\bf 0.14(5)} & {\bf 6/8} & 0.15 & ? & NSA & 0.24(5) & 0.0176(8) & 51/49 & 36(15)\\
& 0.3(7) & 0.06(4) & 25/12 & {\bf  20(20)} & {\bf 0.15(11)} & {\bf 22/11} & 0.71 & ? & NSA & 0.3 f & 0.029(9)& 7/11 &  3(2)\\
 & 0.9(5) & 0.08(2) & 20/12 & {\bf 36(2)} & {\bf 0.30(5)} & {\bf 6/11} & 5 E$-$4 & A$^{\dagger}$ & NSA & 0.3 f & 0.027(11) & 18/14 & 1.5(10)\\\vspace{0.07in}
& 0.32(6) & 0.20(2) & 32/8 & {\bf 25(3)} & {\bf 0.35(4) }& {\bf 7/8} & 6 E$-$4 & A & NSA & 0.40(6) & 0.0171(8) & 73/42 & 38(12)\\

57 & 0.22(5) & 0.09(2) & 9/9 & {\bf 9(5) }& {\bf 0.11(3) } & {\bf 8/8} & 0.49 & ? &  NSA &  0.1 f & 0.0199(6) & 60/52 & 20(3)\\
& 0.8(7) & 0.18(3) & 30/12 & {\bf 23(7)} & {\bf 0.44(12)} & {\bf 16/11} & 8 E$-$3 & ? & NSA & 0.1 f & 0.030(4) & 4/10 & 0.7(5) \\
& 1.1(6) & 0.12(2) & 44(9) & {\bf 41(5)} & {\bf 0.52(9)} & {\bf 11/8} & 1 E$-$3 & A$^{\dagger}$ & $\dots$ & $\dots$& $\dots$ & $\dots$\\\vspace{0.07in}
& 0.45(11) & 0.17(2) & 68/9 & {\bf 36(3)} & {\bf 0.50(6)} & {\bf 13/8} & 4 E$-$4 & A & PO & 0.05(5) & 2.3(3) & 36/41 & 0.62(14)\\

58 & 0.31(18) & 0.09(2) & 10/9 & {\bf 35(13)} & {\bf 0.20(10)} & {\bf 10/8} & 0.22&  ? & BB & 0.1 f & 0.128(8) & 49/35 & 1.2(3)\\
& 0.3(3) & 0.18(4) & 10/9 & {\bf 39(9)} & {\bf 0.47(15)} & {\bf 8/9} & 0.86 & ? & BB & 0.1 f & 0.12(3) & 8/8 & 0.8(6) \\
 & 0.2(3) &0.07(2) & 11/9 & {\bf 24(2)} & {\bf 0.17(4)} & {\bf 4/8} & 3 E$-$3 & ? & BB & 0.21(11) & 0.17(2) & 75/86 & 2.1(4) \\\vspace{0.07in}
& 0.32(8) & 0.20(2) & 88/12 & {\bf 33(3)} &{\bf  0.56(6)} & {\bf 14/11} & 1 E$-$5 & A & BB & 0.20(10) & 0.15(2) & 39/27 & 1.4(3)\\

59 & 0.11(13) & {\bf 0.05(2)} & {\bf 6/12} & 50(20) & 0.21(15) & 7/11 & NA & $-$ & PO & 0.15(9) & 0.86(16) & 50/38 &7.8(10) \\
& 0.14(12) &0.13(3) & 43/12 & {\bf 57.8(5) } & {\bf 1.3(2) } & {\bf 18/12} & 2 E$-$3 & A$^{\ast}$ & PO & 0.3(2) & 1.1(3) & 4/6 & 9.1(15) \\
& 0.20(8) & 0.14(2) & 45/9 & {\bf 27(4)} & {\bf 0.36(15)} & {\bf 8/8} & 3 E$-$4 & A  & PO & 0.12(8) & 0.94(14) & 115/92 & 6.7(8) \\\vspace{0.07in}
& 0.16(6) & 0.15(2) & 37/9  &{\bf 28(4)} & {\bf 0.32(5)} & {\bf 9/8} & 0.18 & ? & PO & 0.17(5) & 1.00(7) & 77/60 & 6.4(5)\\

60 & 0.19(10) & 0.08(2) & 20/12 & {\bf 36(8)} &{\bf 0.28(8)} &{\bf 12/11} & 0.01 & ? &  PO & 0.9(3) & 0.8(2) &42/38 &  8.8(5)\\
& 0.24(14) & 0.17(3) & 14/12 & {\bf 34(7)} &{\bf  1.9(2)} & {\bf 3/11} & 1 E$-$4 & A & PO & 0.9(7) & 1.1(4) & 10/6 & 6.8(12)\\
 & 0.15(15) & 0.08(2) & 18/9 & {\bf 37(9)} & {\bf 0.27} & {\bf 10/8} & 0.03 & ? & PO & 0.9(4) & 0.9(2) & 52/43 & 8.0(10)\\\vspace{0.07in}
& 0.28(6)& 0.16(2) & 61/12 & {\bf 39(4) } & {\bf 0.51(8)} & {\bf 13/11} & 5 E$-$5 & A & PO & 0.9(2) & 0.84(16) & 74/46 & 5.8(3)\\
\noalign{\smallskip}
\hline
\noalign{\smallskip}
\end{tabular}

\end{table*}

\clearpage
\setcounter{table}{1}
 \begin{table*}[!t]   
\renewcommand{\baselinestretch}{.5}
\renewcommand{\tabcolsep}{1mm}
\caption{ continued. }
\begin{tabular}{llllllllllllllll}
\noalign{\smallskip}
\hline
\noalign{\smallskip}
& & \multicolumn{2}{c}{Power Law Fit} & \multicolumn{3}{c}{Broken Power Law Fit}& & &\multicolumn{5}{c}{SED FIT} \\
S & { $F_{\rm var}$} & $\gamma$ & $\chi^2_{\nu}$ & $\nu_{\rm c}$ & $\beta$ & $\chi^2_\nu$& $F$  & Type & Spec& $N_{\rm H}$ & $\Gamma$/kT & $\chi^2_\nu$ &   L\\
\noalign{\smallskip\hrule\smallskip}
61 &0.23(9) & 0.16(2) & 9/9 & {\bf 27(6)} & {\bf 0.31(6)} & {\bf 4/8} & 0.02 & ? & PO & 0.16(6) & 2.4(3) & 35/41 & 3.0(3)\\
& 0.27(14) & 0.12(3) & 24/9 & {\bf 42(7)} & {\bf 0.58(17) } & {\bf 9/8} & 6 E$-$3 & ? & PO & 0.22(13) & 2.2(4) & 4/6 & 2.5(4)\\
 & 0.23(10) & 0.18(2) & 31/9 & {\bf 31(4)} & {\bf 0.44(7)} &{\bf 5/8} & 2 E$-$4 & A & PO & 0.16(6) & 2.1(3) & 101/92 & 2.5(3)\\\vspace{0.07in}
& 0.13(7) & 0.12(2) & 35/9 & {\bf 35(5) } & {\bf 0.32(6)} & {\bf 7/8} & 4 E$-$4 & A$^{\dagger}$ & PO & 0.20(3) & 2.26(13) & 63/72 & 3.8(3)\\

62 &  0.09(7) & 0.07(2) & 15/12 & {\bf 38(11)} & {\bf 0.25(10)} & {\bf 6/11} & 1 E$-$3  & ? & PO & 0.09(3) & 1.42(10) & 56/71 & 8.4(7)\\
& 0.12(10) & 0.16(4) & 20/9 & {\bf 42(6)} & {\bf 0.60(17)} & {\bf 10/8} & 0.02 & ?&PO & 0.08(6) & 1.5(2) & 19/16 &  6.6(8)\\
& 0.08(7) & {\bf 0.12(2)} & {\bf 14/9} & 57(5) & 0.6(3) & 21/8 & NA & ? & PO &  0.11(2) & 1.62(10) & 173/151 & 7.7(5)\\\vspace{0.07in}
& 0.14(3) & 0.18(2) & 42/9 & {\bf 29(3)} & {\bf 0.40(4)} & {\bf 6/8} & 2 E$-$4 & A$^{\dagger}$ & PO & 0.14(2) & 1.70(7) & 125/108 & 6.8(3)\\

63  & $\dots$ &  $\dots$ &  $\dots$ &  $\dots$ &  $\dots$ &  $\dots$ &  $\dots$ &  $\dots$ &  $\dots$ &  $\dots$ &  $\dots$ &  $\dots$ &  $\dots$  \\
& 0.26(16) & 0.22(3) & 23/12 & {\bf 31(7)} & {\bf 0.63(15) }& {\bf 6/11} & 1 E$-$4 & A & PO & 0.22(13) & 3.3(8) & 8/13 & 3.1(6)\\
& $\dots$ &  $\dots$ &  $\dots$ &  $\dots$ &  $\dots$ &  $\dots$ &  $\dots$ &  $\dots$ &  $\dots$ &  $\dots$ &  $\dots$ &  $\dots$ &  $\dots$  \\\vspace{0.07in}& 0.21(2) & 0.17(6) & 30/12 & {\bf 25(4)} & {\bf 0.25(4)} & {\bf 8.3/11} & 2 E$-$4 & A & PO & 0.23(3) & 2.84(16) &  66/65 & 6.4(5)\\
\noalign{\smallskip}
\hline
\noalign{\smallskip}
\end{tabular}
\end{table*}

\end{document}